\documentclass[aps,prx,twocolumn,showpacs,superscriptaddress]{revtex4-2}
\usepackage{graphicx}
\usepackage{amsmath}
\usepackage{amssymb}
\usepackage{colordvi}
\usepackage{mathrsfs}
\usepackage{bm}
\usepackage{verbatim}
\usepackage{dcolumn}
\usepackage{bm}
\usepackage{epsfig}
\usepackage{subfigure}
\usepackage{dsfont}
\usepackage{bbding}

\usepackage[colorlinks,linkcolor=blue,anchorcolor=blue,citecolor=blue]{hyperref}

\begin{document}
	\title{Multipolar Anisotropy in Anomalous Hall Effect from Spin-Group Symmetry Breaking}
	\author{Zheng Liu}
	\affiliation{CAS Key Laboratory of Strongly-Coupled Quantum Matter Physics, and Department of Physics, University of Science and Technology of China, Hefei, Anhui 230026, China}
	\author{Mengjie Wei}
	\affiliation{CAS Key Laboratory of Strongly-Coupled Quantum Matter Physics, and Department of Physics, University of Science and Technology of China, Hefei, Anhui 230026, China}
	\author{Dazhi Hou}
	\affiliation{ICQD, Hefei National Laboratory for Physical Sciences at Microscale, University of Science and Technology of China, Hefei, Anhui 230026, China}
	\author{Yang Gao}
        \email[Correspondence author:~~]{ygao87@ustc.edu.cn}
	\affiliation{CAS Key Laboratory of Strongly-Coupled Quantum Matter Physics, and Department of Physics, University of Science and Technology of China, Hefei, Anhui 230026, China}
        \affiliation{ICQD, Hefei National Laboratory for Physical Sciences at Microscale, University of Science and Technology of China, Hefei, Anhui 230026, China}
        \author{Qian Niu}
	\affiliation{CAS Key Laboratory of Strongly-Coupled Quantum Matter Physics, and Department of Physics, University of Science and Technology of China, Hefei, Anhui 230026, China}

	\date{\today{}}
	
	\begin{abstract}

 Traditional view of the anomalous Hall effect~(AHE) in ferromagnets is that it arises from the magnetization perpendicular to the measurement plane and that there is a linear dependence on the latter.  Underlying such a view is the thinking that the AHE is a time-reversal symmetry breaking phenomenon and can therefore be treated in terms of a power series in the magnetic order. However, this view is squarely challenged by a number of experiments recently, urging for a thorough theoretical investigation on the  fundamental level. We find that for strong magnets, it is more appropriate and fruitful to regard the AHE as a spin-group symmetry breaking phenomenon where the critical parameter is the spin-orbit interaction strength, which involves a much smaller energy scale. In ferromagnets, the spin-orbit coupling breaks the $\infty 2^\prime$ spin rotation symmetry, and the key to characterizing such symmetry breaking is the identification of spin-orbit vectors which transform regularly under spin group operations. Born out of our framework is a rich multi-polar relationship between the anomalous Hall conductivity and the magnetization direction, with each pole being expanded progressively in powers of the spin-orbit coupling strength.  For the leading order contribution, i.e., the dipole, its isotropic part corresponds to the traditional view, and its anisotropic part can lead to the in-plane AHE where the magnetization lies within the measurement plane. Beyond the dipolar one, the octupolar structure offers the leading order source of nonlinearity and hence introduces unique anisotropy where the dipolar structure cannot. The dipolar and octupolar structure offers a unified explanation for the in-plane AHE recently observed in various ferromagnets, and our comprehensive analysis further extends the candidate material systems. Our theory lays the ground for decoding the coupling between various transport and optical phenomena and the magnetic orders.
	\end{abstract}
	
	\maketitle
	
		\section{Introduction}
	As a fundamental phenomenon in solid state physics, the anomalous Hall effect demonstrates the intimate relation between the microscopic band geometry, i.e., the Berry curvature, and the macroscopic order parameter, and hence has attracted intense research interest\cite{SINOVA2004,Nagaosa2006,Yao2004,Xiao2010,Pugh1953,Nagaosa2010}. Experimentally, the Hall resistivity is often explained by the following empirical relation\cite{Pugh1953,Nagaosa2010,Kundt1893,Pugh1930,Pugh1932,Husmann2006,Onose2006,Manyala2004,Zeng2006}:
	\begin{align}\label{eq_emp}
		\rho_{xy}=R_0 H_z+R_s M_z\,.
	\end{align}
	The second term in Eq.~\eqref{eq_emp} essentially describes the structure of the anomalous Hall resistivity in the magnetic-order space: $\rho_{xy}$ depends linearly on the magnetic order and the Hall deflection plane is perpendicular to the direction of the magnetic order. It is this structure that leads to the measuring geometry of the anomalous Hall effect in experiments.
	
	However, in recent years, the uniqueness of this structure has been seriously challenged both theoretically and experimentally\cite{Chen2014,Nakatsuji2015,Nayak2016,Ikhlas2017,Zhang2017,Yang2017,Liu2018,Zhao2019,Chen2021,Liu2007,Roman2009,Tan2021,Cao2023,Zhou2022,Wang2024,Battilomo2021,Wang2023,Peng2024}. Studies in ${\rm Mn_3X}$~(X=Ir, Sn, Ge, Pt, Ge, Rh) show that besides the ferromagnetic order, the anomalous Hall effect can also be tied with noncollinear antiferromagnetic orders with no net spin magnetization~(i.e., $\bm M=0$), as long as permitted by the magnetic point group symmetry\cite{Chen2014,Nakatsuji2015,Nayak2016,Ikhlas2017,Zhang2017,Yang2017,Liu2018,Zhao2019,Chen2021}. Recently, the developments of the spin-group theory further extends this correlation in the absence of spin-orbit coupling\cite{Liu2022,ifmmodeSelseSfimejkal2022,Chen2024,Xiao2024,Jiang2023,McClarty2024}. Moreover, even with ferromagnetic orders, the quantum geometry can show striking anisotropy in crystals with sufficiently low point-group symmetry, such that the magnetization lies within the Hall deflection plane, which is referred to as the in-plane anomalous Hall effect\cite{Liu2007,Roman2009,Tan2021,Cao2023,Zhou2022}. There even exists complicated patterns in the anomalous Hall signal as the magnetization varies in the Hall-deflection plane\cite{Battilomo2021,Wang2023,Wang2024,Peng2024}. These advancements naturally raise a fundamental question: what is the complete and accurate understanding of the structure of the anomalous Hall resistivity or conductivity in the configuration space of magnetic orders?
	
	Traditionally, the anomalous Hall effect is treated as a time-reversal symmetry breaking effect. Previous understanding such as the empirical law then suggests the Taylor expansion of the anomalous Hall conductivity with respect to the magnetic order\cite{Birss1964,Hurd1974,Petukhov1998}. The Hall-type conductivity $\sigma_i^H=\frac{1}{2}\epsilon_{ijk}\sigma_{jk}$ is generally subject to the Onsager's reciprocal relation\cite{Onsager1931}, which dictates that $\bm \sigma^H(\bm M)=-\bm \sigma^H(-\bm M)$. Consequently, the Taylor expansion of $\bm \sigma^H$ with respect to $\bm M$ should only contain odd-order terms. In such framework, the leading order term is linear with a tensorial coefficient, whose isotropic and anisotropic part account for the empirical law and the in-plane anomalous Hall effect, respectively. The anomalous Hall effect in noncollinear antiferromagnets can also be explained by replacing $\bm M$ with noncollinear antiferromagnetic order\cite{Suzuki2017,Suzuki2019}. 
	
	However, such expansion is inherently flawed as it requires that the exchange energy associated with the magnetic order is much smaller than all the other relevant energy scales, a condition that is rarely met in ferromagnetic materials\cite{Stoehr2006,Cowan1981,Novak2001,Khomskii2021,Dunn1961,Yuan2017,Stamokostas2018,Tanaka2008,Naito2010,Herman1963}. In fact, the electron-electron interaction that gives rise to the magnetic order, including the Hubbard energy and  the $s$-$d$ exchange coupling, is usually comparable with or even larger than the hopping strength. This leads to an irreconcilable conflict: the nonlinear terms in the power series expansion of $\bm \sigma^H$ over $\bm M$ should dominate over the linear term, which has rarely been observed.  The mathematical and experimental disapproval of performing Taylor expansion over the magnetic order suggests that to understand the structure of the anomalous Hall effect, even the simplest one as given by the empirical law in Eq.~\eqref{eq_emp}, a completely different framework is required.
 
    Moreover, previous theory of the empirical law and the in-plane anomalous Hall effect hides the essential role of the spin-orbit coupling. The spin-orbit coupling usually bridges the chiral orbital motion and the time-reversal symmetry breaking due to the spin order, and is necessary for a nonzero anomalous Hall effect in ferromagnets and coplanar antiferromagnets\cite{Yao2004,GosalbezMartinez2015,Zhang2011,Guo2014,Suzuki2017}. Although recent spin-group theory suggests that in materials with non-coplanar spin orders, the anomalous Hall effect can appear without spin-orbit coupling\cite{Chen2024}, in practical scenarios where the variation of the anomalous Hall signal with the rotation of the total spin order is needed, the spin-orbit coupling is still indispensable. Previously, there are persistent interest in identifying the detailed relation between the anomalous Hall effect and the spin-orbit coupling\cite{Yao2004,Zhang2011,Guo2014,Zhang2016,Zhang2011a,Fan2015,He2012}. However, a comprehensive and accurate framework is still lacking. 	
    
	In this work, we provide such a framework in ferromagnets by establishing the anomalous Hall effect as a spin-group symmetry breaking phenomenon. This is achieved by treating the spin-orbit coupling as a perturbation, which is well justified as the strength of the spin-orbit coupling is usually the smallest energy scale in all the relevant energy scales for the anomalous Hall effect\cite{Stoehr2006,Cowan1981,Novak2001,Khomskii2021,Dunn1961,Yuan2017,Stamokostas2018,Tanaka2008,Naito2010,Herman1963}. We show that based on its SU(2) gauge structure, the spin-orbit coupling can be parameterized by a set of spin-orbit vectors $\lambda_i^a$ with $i=x,y,z$ for spatial directions and $a=1,2,3$ acting as a color index. The order-by-order perturbation of the spin-orbit coupling to $\bm \sigma^H$ then reduces to a power series of $\lambda_i^a$, whose coefficients at each order contain two sets of indices, i.e., the spatial and color indices. For the Hamiltonian without spin-orbit coupling in ferromagnets, the spin group symmetry is the direct product of the spin-only group $\infty 2^\prime$ and the crystalline point group\cite{Liu2022,Chen2023}. This then imposes the point group symmetry in color space obeyed by $\bm \sigma^H$ to be $D_\infty$. As an important consequence, the power series of $\bm \sigma^H$ with respect to $\lambda_i^a$ transforms into  a multipolar expansion of $\bm \sigma^H$ in the magnetic-order space,  providing a clear understanding of the structure of $\bm \sigma^H$ and hence a more accurate and completer version of Eq.~\eqref{eq_emp} . The poles of $\bm \sigma^H$ in the magnetic-order space characterize the distribution of the Berry curvature in the magnetic-order space, in sharp contrast to the Berry curvature multipoles in the momentum space\cite{Nagaosa2010,Sodemann2015,Sankar2024,Zhang2023} and the multipoles of the spin order in real space\cite{Suzuki2017,Suzuki2019,Ederer2007,Hayami2022,Yatsushiro2021,Yanagi2023,Bhowal2024}. 

 This multipolar expansion is completely different from the previous Taylor expansion over the magnetic order $\hat{\bm M}$. First, it involves explicitly only the direction of $\hat{\bm M}$, hence unaffected by the largeness of the magnitude of $\bm M$. Secondly, the spin-orbit coupling naturally and progressively enters in the various poles, highlighting its elemental role and guaranteeing the convergence of the expansion. 

The multipolar expansion in the magnetic-order space has a profound implication in the anomalous Hall effect.  Theoretically, it decodes the coupling between the geometric object such as the Berry curvature and the fundamental lattice structure. Experimentally, it replaces Eq.~\ref{eq_emp} and serves as  a more accurate and complete law for explaining relevant data, based on which the nonliearity in the magnetic order enters naturally. By performing a comprehensive analysis of the dipolar and octupolar structures of $\bm \sigma^H$ in the magnetic-order space, we demonstrate the resulting unique anistropy and provide exact guiding rules for observing the in-plane anomalous Hall effect.
	
	Our paper is organized as follows. In Sec. II, we provide the general framework for the multipolar expansion of $\bm \sigma^H$. This is done by introducing the spin-orbit vectors, treating the spin-orbit coupling as a perturbation, and deriving the point-group symmetry of $\bm \sigma^H$ in the color space. In Sec. III, we provide a comprehensive analysis of the dipolar structure of $\bm \sigma^H$ in the magnetic-order space. We show that the symmetric and toroidal dipole contributes distinctly to the conventional part and the in-plane part of the anomalous Hall effect. Using symmetry analysis, we are able to provide guiding rules for observing the latter. We also demonstrate the dipolar structure in ferromagnetic $\delta$-MnGa using first-principles calculations. In Sec. IV, we analyze the octupolar structure, which provides the leading-order nonlinearity in $\bm \sigma^H$. We show that although generally the octupolar structure mixes with the dipolar one, there are four cases of the in-plane anomalous Hall effect where only the octupolar structure contributes and the dipolar one does not. The octupolar structure then greatly extends the possibility of observing the in-plane anomalous Hall effect. We demonstrate the octupolar structure in fcc Ni and magnetic Weyl semimetal  Co$_3$Sn$_2$S$_2$ using first-principles calculations.  In Sec. VI, we summarize the implication of our theory and discuss its generalization in antiferromagnets and potential applicability in studying other transport and optical phenomena.
 
	\begin{table*}[t]
		\caption{The Hall-type current due to the multipole of the Berry curvature in the momentum and magnetic-order space.}
		\begin{ruledtabular}
			\begin{tabular}{ccc}
				& in momentum space & in magnetic-order space  \\ \hline    
				Monopole $c_i$     &  $\bm J=\bm E\times \bm c$~\cite{Nagaosa2010} &    Zero\\
				Dipole $p_{ij}$     &  $\bm J= \tau \bm E\times \bm p\cdot \bm E$~\cite{Sodemann2015}    & $\bm J=\bm E\times \bm p \cdot \hat{\bm M}$ \\
				Quadrupole $q_{ijk}$   &  $\bm J= \tau^2 \bm E\times \bm q\cdot \bm E\bm E$~\cite{Sankar2024,Zhang2023} & Zero\\
				Octupole $o_{ijk\ell}$  & $\bm J= \tau^3 \bm E\times \bm o\cdot \bm E\bm E\bm E$~\cite{Zhang2023} &    $\bm J=\bm E\times \bm o \cdot \hat{\bm M}\hat{\bm M}\hat{\bm M}$ 
			\end{tabular}
		\end{ruledtabular}
		\label{table-0}
	\end{table*}	
 
	\section{The multipolar expansion of the anomalous Hall conductivity}
	\subsection{The multipole series}
	The fundamental constraint on the anomalous Hall conductivity is the Onsager's reciprocal relation\cite{Onsager1931}, which, in ferromagnetic materials, reads as $\bm \sigma^H(\bm M)=-\bm \sigma^H(-\bm M)$. The generalization to antiferromagnets will be discussed in the Discussion section. To further explore the fine structure of $\bm \sigma^H$ in the magnetization space, the strength of relevant configurations needs to be quantified. As explained previously, the Taylor expansion of $\bm \sigma^H$ with respect to $\bm M$ is generally invalid. Therefore, we take another route and consider the multipolar structure of $\bm \sigma^H$. Due to the Onsager's reciprocal relation, any even-order pole of $\bm \sigma^H$ vanishes identically.

	The first nonzero pole is then the dipole of $\bm \sigma^H$, which can be defined as follows
	\begin{align}\label{eq_dip}
		p_{ij}=\frac{3}{4\pi} \int \sin\theta d\theta d\phi \sigma_i^H(\theta,\phi) \hat{M}_j\,,
	\end{align}
	where $\hat{\bm M}$ is a unit vector along $\bm M$ with the polar angle $\theta$ and azimuthal angle $\phi$. Since both $\bm \sigma^H$ and $\hat{\bm M}$ transform as an axial vector, $p_{ij}$ should transform as a rank-2 tensor. Therefore, for a point group symmetry represented by an orthogonal matrix $S$, the following constraint is imposed
	\begin{align}
		p=SpS^T\,.
	\end{align}
	
	The dipole $p_{ij}$ contributes to $\bm \sigma^H$ linearly as follows
	\begin{align}\label{eq_dip_sig}
		\bm \sigma^{H,1}=p_{ij}\hat{M}_j \hat{e}_i\,.
	\end{align}
	Here and hereafter, repeated indices are summed unless otherwise specified.
	
	
	The next nonzero order is the octupole of $\bm \sigma^H$, defined as follows
	\begin{align}\label{eq_oct}\
		o_{ijk\ell}=\frac{7}{8\pi}\int \sin\theta d\theta d\phi &\sigma_i^H(\theta,\phi)(5\hat{M}_j\hat{M}_k\hat{M}_\ell -\hat{M}_j \delta_{k\ell}\notag\\
		&-\hat{M}_k\delta_{j\ell}-\hat{M}_\ell \delta_{jk})\,.
	\end{align}
	Its contribution to the anomalous Hall conductivity is cubic:
	\begin{align}\label{eq_octsym}
		\bm \sigma^{H,3}=o_{ijk\ell}\hat{M}_j \hat{M}_k \hat{M}_\ell \hat{e}_i\,.
	\end{align}

	The above octupole has several essential properties. First, it is straightword to find that $o_{ijk\ell}$ transform as a rank-4 tensor, which is also symmetric with respect to the last three indices. Secondly, $o_{ijk\ell}$ is traceless, i.e., $\sum_k o_{ijkk}=\sum_j o_{ijj\ell}=\sum_j o_{ijkj}=0$. From the first property, there are 10 different non-equivalent choices of the last three indices in three dimensions. The traceless property further reduces this number to 7. It is important to notice that this number agrees with the number of orbitals with angular momentum $\ell=3$. In fact, the factor in the bracket of Eq.~\eqref{eq_oct} can be put into linear combinations of the spherical harmonics $Y_{\ell m}(\theta,\phi)$ with $\ell=3$. Therefore, the dipole and octupole contribution to the anomalous Hall conductivity is independent: $\bm \sigma^{H,1}$ does not have an octupole element and $\bm \sigma^{H,3}$ does not have a dipole element.
	
	The above discussion can be easily generalized to all the odd-order poles. In terms of them, the anomalous Hall conductivity reads as
	\begin{align}\label{eq_mexp}
		\bm \sigma^H=\sum_{a\in N} \bm \sigma^{H,2a+1}\,.
	\end{align}
	However, it is important to emphasize that by far the above expression is not yet a series expansion. To make it so, we need to further prove that as the order of poles increases, the strength decays fast enough. 
	
	It is important to point out that the multipoles as discussed above is obtained by ignoring the magnetic anisotropy energy $E_M$. Therefore, they solely reflects the multipolar structure of the quantum geometry~(i.e., Berry curvature) in the magnetic-order space. In general, however, the total energy can be decomposed as $E=E_0+E_M$ and the multipoles should also be a function of $E_M$ and hence the energy anisotropy also manifests in the anomalous Hall effect. The relevant effect can in fact be incorporated in our framework as shown later.
	
	Before ending this section, we would like to point out the difference between the multipole in our theory and those in previous studies. First, in antiferromagnets, the spin order in a unit cell can have a nontrivial pattern, which can be characterized by a series of real-space multipole, similar to the multipole of charge and current in electromagnetic theory\cite{Suzuki2017,Suzuki2019,Ederer2007,Hayami2022,Yatsushiro2021,Yanagi2023,Bhowal2024}. This type of multipole usually act as the primal order parameter. Secondly, since the Berry curvature is generally inhomogeneous in momentum space, its momentum-space multipole has been extensively studied recently, which provides valuable interpretations to the power-law nonlinearity of the Hall current with respect to the electric field\cite{Nagaosa2010,Sodemann2015,Sankar2024,Zhang2023}. Here in our theory, we recognize a series of magnetization-space multipoles, and they reveal the multipolar nonlinearity and hence anisotropy of the anomalous Hall conductivity in the magnetic order space as discussed later. As both the momentum-space multipole and the magnetization-space multipole correspond to certain nonlinearity in the Hall-type current, we summarize their difference in Table.~\ref{table-0}.
	
	\subsection{The fine structure of the spin-orbit coupling}
	
	The spin-orbit coupling is indispensable in understanding many phenomena related to magnetism. For example, without the spin-orbit coupling, the exchange coupling between spins would be isotropic with respect to different global spin orientations. The spin-orbit coupling then link the spin orientation with the crystal field anisotropy and hence breaks this continuous symmetry in the spin space, leading to the uni-axial anisotropy\cite{Coey2010} and the Dyashinskii-Moriya interaction\cite{Dzyaloshinsky1958,Moriya1960}. For the anomalous Hall effect, the spin-orbit coupling makes the orbital motion sense the time-reversal-symmetry breaking in the spin degree of freedom. In ferromagnets, the anomalous Hall effect always vanishes without the spin-orbit coupling\cite{GosalbezMartinez2015}.
	
	Although important, the spin-orbit coupling is usually quite small, as can be seen from its expression. It is a relativistic effect and arises as the Dirac Hamiltonian is projected onto the upper branches. In ferromagnetic materials, the Schrodinger Hamiltonian can be generally put in the following form:
	\begin{align}\label{eq_ham0}
		\hat{H}=\hat{H}_0+\hat{H}_{int}+\hat{H}_{soc}\,,
	\end{align}
	where $\hat{H}_0=\frac{\hat{\bm p}^2}{2m}+U(\bm r)$ is independent of spin, $\hat{H}_{int}=J(\bm r)\hat{\bm M}\cdot \bm \sigma$ is the exchange coupling with $\hat{\bm M}$ representing the direction of the ferromagnetic order, and 
	\begin{align}\label{eq_soc0}
		\hat{H}_{soc}=\frac{\hbar}{m^2c^2}\hat{\bm O}\cdot \bm \sigma
	\end{align}
	is the spin-orbit coupling with $\hat{\bm O}=\bm \nabla U(\bm r)\times \hat{\bm p}$. In the case of central potential, $\hat{\bm O}$ is proportional to the orbital angular momentum. Due to the appearance of the speed of light, $\hat{H}_{soc}$ should be small compared with $\hat{H}_0+\hat{H}_{int}$.
	
	In practice, materials consisting of transition metal elements with partially filled $d$-orbitals include a large group of ferromagnets. In them, the spin-orbit coupling usually competes with the following three energy scales: the crystal field splitting, the electron-electron interaction, the hopping strength from the kinetic energy. Generally speaking, when one goes from 3$d$ to 4$d$ to 5$d$ transition metals, the electron-electron interaction decreases, the crystal field splitting and the spin-orbit coupling increase, and the typical band width due to hopping does not change much. For 3$d$ and 4$d$ transition metal elements, the spin-orbit coupling is on the order of 0.01-0.1~eV, while the other three are on the order of eV\cite{Stoehr2006,Cowan1981,Novak2001,Dunn1961,Khomskii2021,Yuan2017,Tanaka2008,Naito2010,Herman1963,Stamokostas2018}. We will thus focus on ferromagnetism originating from 3$d$ and 4$d$ transition metal elements, for which the spin-orbit coupling can be safely treated as perturbation. 
	
	To study the role of the spin-orbit coupling, previous literatures focus on its strength, which can be easily tuned by varying the speed of light based on Eq.~\eqref{eq_soc0}. Using this method, it is shown that the anomalous Hall effect depends almost linearly on the spin-orbit coupling in bcc iron\cite{Yao2004}. There is also finer method which differentiates the spin-conserving part from the spin-flip part in the spin-orbit coupling, as the former alone preserves the spin component in the direction parallel with the ferromagnetic moment\cite{Zhang2011,Guo2014}. By keeping each part of the spin-orbit coupling alone and varying the speed of light, one obtains a finer understanding about the dependence on the spin-orbit coupling.

    \begin{figure}
		\includegraphics[width=7.0cm,angle=0]{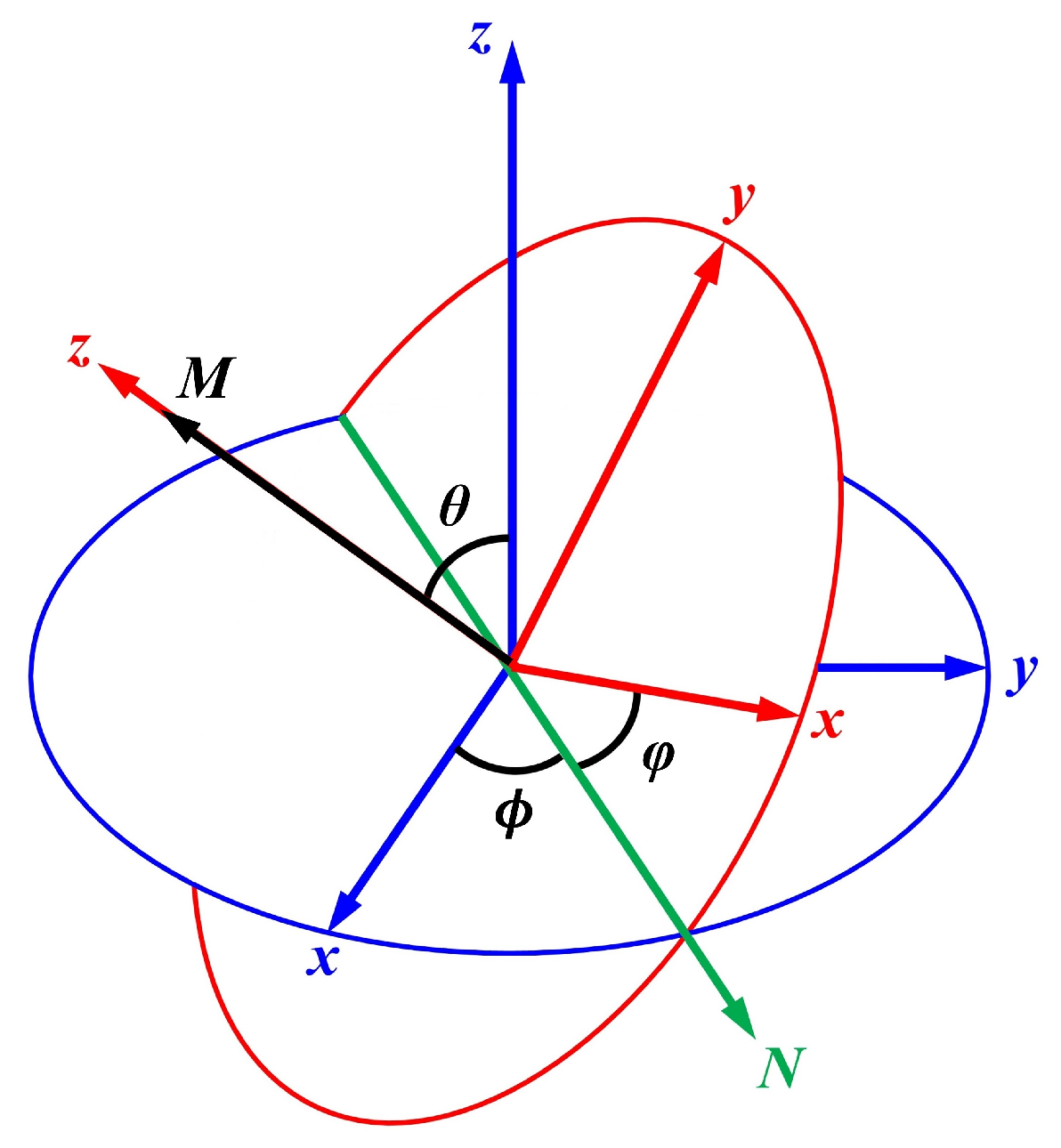}
		\caption{The relative orientation of the spin frame and crystal frame.}
		\label{Fig_euler}
    \end{figure}

	Besides the strength, the spin-orbit coupling has a richer structure related to the magnetic order hidden inside its SU(2) form. Upon changing $\hat{\bm M}$, the preferable spin orientation in $\hat{H}_{int}$ changes. To reflect this property, we allow the crystal and the spin to have different coordinate systems, and the latter one is defined so that $\bm M$ is always along the $z$ direction.  The relative orientation of the spin frame with respect to the crystal frame can then be parameterized by the three Euler angles as $\phi$, $\theta$, and $\varphi$, as shown in Fig.~\ref{Fig_euler}.  Then the spin-orbit coupling has the following form
	\begin{align}\label{eq_socvec}
		\hat{H}_{soc}=\frac{\hbar}{m^2c^2}\sum_{a=1}^3 (\bm \ell^a\cdot\hat{\bm O})\sigma_a\,.
	\end{align}
	Compared with Eq.~\eqref{eq_soc0}, Eq.~\eqref{eq_socvec} involves a set of three vectors $\bm \ell^a$. These $\bm \ell^a$ can be expressed using the three Euler angles, and in the case that $\varphi=0$, we have
	\begin{align}\label{eq_socvec2}
		\bm \ell^1&=(\cos\phi,\sin\phi,0)\notag\\
		\bm \ell^2&=(-\cos\theta\sin\phi,\cos\theta \cos\phi,\sin\theta)\notag\\
		\bm \ell^3&=(\sin\theta\sin\phi,-\sin\theta\cos\phi,\cos\theta)\,.
	\end{align}
	We will refer to $\bm \ell^a$ as the spin-orbit vector and the index $a$ as the color index. From the definition in Eq.~\eqref{eq_socvec}, it is straightforward to find that $\bm \ell^3=\hat{\bm m}$ and $\bm \ell^1\times \bm \ell^2=\bm \ell^3$. We emphasize that this just provides the value for these spin-orbit vectors and does not necessarily imply their transformation properties.
	
	The spin-orbit vector and the color degree of freedom full reveals the structure of the spin-orbit coupling. They describe how the deviation of the spin frame from the crystal frame affects the Hamiltonian and hence various phenomena. Moreover, the spin-conserving part and the spin-flip part can be easily defined in this language: $\bm \ell^3$ characterizes the spin-conserving part and $\bm \ell^1$ and $\bm \ell^2$ characterize the spin-flip part.
	
	The three spin-orbit vectors are orthonormalized in both the spatial and color degree of freedom.  It is straightforward to find that $\ell_i^a\ell_i^b=\delta_{ab}$ since $\bm \ell^a$ should span the coordinate system in the spin space. Moreover, originally in Eq.~\eqref{eq_soc0}, the three spin-orbit vectors are $\ell^{1,o}=(1,0,0)$, $\ell^{2,o}=(0,1,0)$, and $\ell^{3,o}=(0,0,1)$. Therefore, $\ell_i^{a,o}\ell_j^{a,o}=\delta_{ij}$. Since the inner product is conserved under the orthogonal transformation, we then reach the conclusion that under a general orientation, $\ell_i^a\ell_j^a=\delta_{ij}$.
	
	\subsection{Spin-group analysis and the point-group symmetry in the color space}
	
	With the help of the spin-orbit vector, the perturbation of the spin-orbit coupling can be efficiently studied. The effect of the spin-orbit coupling on the wavefunction can be derived order by order following the standard perturbation theory. For the anomalous Hall effect, this necessarily leads to an expansion of the anomalous Hall conductivity with respect to the spin-orbit coupling, which consists of terms with different powers of the factor $\hbar/m^2c^2$. However, since the spin-orbit coupling is always linear in the spin-orbit vector, such an expansion can be alternatively expressed as a power series over the spin-orbit vector, i.e.,
	\begin{align}\label{eq_exp1}
		\sigma_i^H=\alpha_{ij}^{\;a}\ell_j^a+\beta_{ijk}^{\; ab} \ell_j^a\ell_k^b+\cdots
	\end{align}
	We emphasize that the $n$-th term in the power series is also of $n$-th order in the strength of the spin-orbit coupling.
	
	The structure of the anomalous Hall conductivity is determined by the nonzero elements of various coefficients. We note that these coefficients have two different sets of indices, one in the color space and the other in the real space. Generally speaking, the coefficients are subject to the spin group symmetry of the Hamiltonian without the spin-orbit coupling. A generic spin group operation can be decomposed into two parts: the spatial part $\mathcal{T}_o$ and the spin-only part $\mathcal{T}_s$. The former acts on $\hat{\bm O}$ and the latter on $\hat{\bm \sigma}$. With the matrix representation $M_o$ for $\hat{\bm O}$ and $M_s$ for $\mathcal{T}_s$, it is straightforward to derive that under the operation $\mathcal{T}_o \mathcal{T}_s$, the spin-orbit vector transform as 
 \begin{align}
     \ell_i^a\rightarrow (M_o)_{ji}\ell_j^b (M_s)_{ba}\,.
 \end{align}
 
 In ferromagnets, this spin group is the direction product of the spin-only group $\infty 2^\prime$ and the crystalline point group\cite{Liu2022,Chen2023}. This simple structure implies that the transformation of the spin-orbit vector in these two spaces can be decoupled. As a result, the real-space elements of the coefficients in Eq.~\eqref{eq_exp1} are subject to the point-group symmetry of the crystal, without counting the magnetic order.
	
	We now analyze the constraint in the color degree of freedom. In ferromagnets, the spin-only group has two generators: $TC_{2x}^s$ and $C_{\varphi z}^s$. Here the upper index $s$ means that the rotation only acts on the spin and $\varphi$ is an arbitrary angle. $\hat{H}_{soc}$ generally does not commute with these operations. Under $TC_{2x}^s$, $\sigma_1$ and $\hat{\bm O}$ flip sign and $\sigma_2$ and $\sigma_3$ stay the same. By viewing $\hat{H}$ as a function of the spin-orbit vector, this change in $\hat{H}_{soc}$ is equivalent to the following transformation in the color space: $\hat{H}(\bm \ell^1,\bm \ell^2,\bm \ell^3)\rightarrow \hat{H}(\bm \ell^1,-\bm \ell^2,-\bm \ell^3)$. As $TC_{2x}^s$ necessarily flips the sign of the Berry curvature, we then obtain 
	\begin{align}\label{eq_const1}
		\bm \sigma^H(\bm \ell^1,\bm \ell^2,\bm \ell^3)=-\bm \sigma^H(\bm \ell^1,-\bm \ell^2,-\bm \ell^3)\,.
	\end{align}
	The symmetry $C_{\varphi z}^s$ originates from the fact that the third Euler angle $\varphi$ is a gauge freedom: in setting up the spin frame, only the $z$ direction is well defined, i.e., the same as $ \hat{\bm m}$; the $x$ and $y$ axis can rotate arbitrarily about $\hat{\bm m}$. Under $C_{\varphi z}^s$, $\sigma_1$ and $\sigma_2$ will mix. This is equivalent to a $C_{\varphi z}$ rotation of the spin-orbit vector in the color space: $\bm \ell^1 \rightarrow \bm \ell^1\cos\varphi+\bm \ell^2 \sin\varphi$ and $\bm \ell^2 \rightarrow -\bm \ell^1\sin\varphi+\bm \ell^2 \cos\varphi$. Since $C_{\varphi z}^s$ should not change the anomalous Hall conductivity, we have
	\begin{widetext}
		\begin{align}\label{eq_const3}
			\bm \sigma^H(\bm \ell^1,\bm \ell^2,\bm \ell^3)=\bm \sigma^H(\bm \ell^1\cos\varphi+\bm \ell^2 \sin\varphi,-\bm \ell^1\sin\varphi+\bm \ell^2 \cos\varphi,\bm \ell^3)
		\end{align}\,.
	\end{widetext}
	The above constraints from Eq.~\eqref{eq_const1} to \eqref{eq_const3} can be put in a compact form: we can view $\bm \sigma^H$ as an element in the color space with $a=3$; with this assignment, it is subject to the $D_\infty$ point group symmetry in the color degree of freedom. This then fully determines the point-group symmetry of the coefficients in Eq.~\eqref{eq_exp1} in the color space.
	
	Finally, as discussed previously, the magnetic anisotropy energy $E_M$ may also contribute to the anomalous Hall conductivity. Using the same recipe, we can obtain the order-by-order correction to the ground state energy due to the spin-orbit coupling. This will reduce to a power series over the spin-orbit vector as follows:
	\begin{align}\label{eq_dE}
		\delta E&= A_i^a \ell_i^a+B_{ij}^{ab} \ell_i^a\ell_j^b+\cdots\,.
	\end{align}
	Similar to the expansion in Eq.~\eqref{eq_exp1}, the real-space elements of the coefficients are subject to the point-group symmetry of the crystal without counting the magnetic order. Since the energy does not change with the time-reversal operation, the spin-group analysis shows that the point-group symmetry in the color space is $C_{\infty v}$.
	Later we will prove that this energy correction is just the magnetic anisotropy energy $E_M=\delta E$.
	
	\subsection{Multipolar expansion}
	The point-group symmetry in the color space dictates that the power series expansion in Eq.~\eqref{eq_exp1} coincides with the multipolar series in Eq.~\eqref{eq_mexp}, making the latter a well-defined multipolar expansion. To see this, we analyze Eq.~\eqref{eq_exp1} order by order. We first note that the  $D_\infty$ symmetry forbids the zero-th order term in Eq.~\eqref{eq_exp1}, consistent with the fact that the anomalous Hall effect in ferromagnets requires the spin-orbit coupling.
	
	At first order, the response coefficient $\alpha_{ij}^{\; a}$ is actually $\alpha_{ij}^{ba}$ with $b=3$ based on the above assignment. The $D_\infty$ symmetry then  suggests that the only nonzero relevant component is $\alpha_{ij}^{33}$. We then omit the upper indice and the corresponding contribution to the anomalous Hall conductivity reads as
	\begin{align}\label{eq_1st}
		\sigma_i^{H,(1)}=\alpha_{ij} \ell_j^3=\alpha_{ij} \hat{ M}_j\,.
	\end{align}
	The last equality holds as $\bm \ell^3$ always traces the direction of $\bm M$.
	
	At second order, the coefficient $\beta_{ijk}^{\; ab}$ is $\beta_{ijk}^{cab}$ with $c=3$. Due to the $D_\infty$ symmetry, the nonzero relevant elements satisfy $\beta_{ijk}^{312}=-\beta_{ijk}^{321}$. We then omit the upper indices of $\beta_{ijk}^{312}$ and the corresponding anomalous Hall conductivity reads as
	\begin{align}\label{eq_2nd}
		\sigma_i^{H,(2)}=\beta_{ijk} (\ell_j^1\ell_k^2-\ell_j^2\ell_k^1)=\tilde{\beta}_{ij} \hat{M}_j\,,
	\end{align}
	where $\tilde{\beta}_{ij}=\epsilon_{jj_1j_2}\beta_{ij_1j_2}$. In the second equality, we use the identity $\bm \ell^1\times \bm \ell^2=\hat{\bm M}$. By comparing Eq.~\eqref{eq_1st} and \eqref{eq_2nd} with Eq.~\eqref{eq_dip_sig}, we immediately identify that $p_{ij}=\alpha_{ij}+\tilde{\beta}_{ij}$. This is the leading order contribution to the dipole of the anomalous Hall conductivity. From the definition of $\alpha_{ij}$ and $\tilde{\beta}_{ij}$, it is then clear that the dipole is at least first order in the spin-conversing part and second order in the spin-flipping part of the spin-orbit coupling, consistent with previous studies.
	
	At third order, the coefficient $\gamma_{ijk_1k_2}^{\;bcd}$ is $\gamma_{ijk_1k_2}^{abcd}$ with $a=3$. Due to the $D_\infty$ symmetry, there are two types of nonzero elements: $\gamma_{ijk_1k_2}^{3333}$ and $\gamma_{ijk_1k_2}^{3113}=\gamma_{ijk_1k_2}^{3223}$. We shall note that $\gamma_{ijk_1k_2}^{abcd}$ is symmetric with respect to the last three pairs of indices, i.e., the element does not change if $(b,j)$, $(c,k_1)$ and $(d,k_2)$ are interchanged with each other. Therefore, $\gamma_{ijk_1k_2}^{3113}$ has two equivalent forms that are implied but not listed explicitly. All the other elements are zero. Moreover, from the symmetry requirement, we find that $ \gamma_{ijk_1k_2}^{3113}$ couples to $\ell_j^1\ell_{k_1}^1+\ell_j^2\ell_{k_1}^2$, which is $\delta_{jk_1}-\ell_j^3\ell_{k_1}^3$ using the orthonormal condition. We can then connect $ \gamma_{ijk_1k_2}^{3113}$ with the direction of $\bm M$. 
	We re-label $\gamma_{ijk_1k_2}^{3333}$ with  $\gamma_{ijk_1k_2}^{1}$ and $\gamma_{ijk_1k_2}^{3113}$ with $\gamma_{ijk_1k_2}^{2}$. The contribution to the anomalous Hall conductivity is then
	\begin{align}\label{eq_3rd}
		\sigma_i^{H,(3)}=&\frac{3}{5}(\gamma_{ijjk}^1+2\gamma_{ijjk}^2)\hat{M}_k+\tilde{\gamma}_{ijk_1k_2} \hat{M}_j \hat{M}_{k_1} \hat{M}_{k_2}\,,
	\end{align}
	where $\tilde{\gamma}_{ijk_1k_2}=\tilde{\gamma}_{ijk_1k_2}^1-3\tilde{\gamma}_{ijk_1k_2}^2$ is traceless,
	\begin{align}
		\tilde{\gamma}_{ijk_1k_2}^1=\gamma_{ijk_1k_2}^1-\frac{1}{5}[\gamma_{ijk^\prime k^\prime} \delta_{k_1k_2}+(j\leftrightarrow k_1)+(j\leftrightarrow k_2)]\,,
	\end{align}
	and $\tilde{\gamma}_{ijk_1k_2}^2$ is defined in the same way from $\gamma_{ijk_1k_2}^2$. We can now compare Eq.~\eqref{eq_3rd} with Eq.~\eqref{eq_mexp}, and find that $p_{ij}=\alpha_{ij}+\tilde{\beta}_{ij}+\frac{3}{5}(\gamma_{ijjk}^1+2\gamma_{ijjk}^2)$ and $o_{ijk_1k_2}=\tilde{\gamma}_{ijk_1k_2}$.

	\begin{table*}[t]
		\caption{The dipolar origin of the in-plane anomalous Hall effect and its symmetry requirement.}
		\begin{ruledtabular}
			\begin{tabular}{ccc}
				Type  & In-plane anomalous Hall effect & Point group  \\ \hline    
				Isotropic dipole    & \XSolid    & $T$, $T_d$, $T_h$, $O$, $O_h$ \\
				Symmetric dipole   & \Checkmark & $C_{2v}$, $C_{3v}$, $C_{4v}$, $C_{6v}$, $D_2$, $D_3$, $D_4$, $D_6$,\\
				&            &   $D_{2h}$, $D_{3h}$, $D_{4h}$, $D_{6h}$, $D_{2d}$, $D_{3d}$\\
				Symmetric and toroidal dipole  & \Checkmark &   $C_1$, $C_2$, $C_3$, $C_4$, $C_6$, $C_{1h}$, $C_{2h}$, $C_{3h}$, \\
				&            &  $C_{4h}$, $C_{6h}$,$S_2$, $S_4$, $S_6$ \\
			\end{tabular}
		\end{ruledtabular}
		\label{table-1}
	\end{table*}
	
	Two comments are in order. First, since the octupole contains both $\gamma^1$ and $\gamma^2$, the leading order contribution is at least at third order in the spin-conserving part of the spin-orbit coupling, or at first order in the spin-conserving part and second order in the spin-flipping part. This is different from the dipole case, where the spin-flipping part is always involved at higher order. Secondly, by comparing the dipole and octupole, we find that the strength of the latter is indeed at least one-order-smaller than the former. This then justifies Eq.~\eqref{eq_mexp} up to the octupole order.
	
	The question is still left whether purely spin-flipping part can contribute to the octupole just as in the dipole case. To answer this, we consider the fourth order in the expansion and the coefficient $\delta_{ijk_1k_2k_3}^{\;bcde}$ is $\delta_{ijk_1k_2k_3}^{abcde}$ with $a=3$. The $D_\infty$ symmetry and the fact that $\delta$ is symmetric with respect to the last four pairs of indices then enforce the following constraints on $\delta$: $bcde$ can only take $1$ or $2$; $\delta$ is antisymmetric with respect to one pair of indices among $bcde$ and diagonal and isotropic with respect the other pair. Since $\bm \ell^e\times \bm \ell^f=\epsilon_{def} \bm \ell^d$, we can then define $(\delta^\prime)_{ijkk_1}^{abcd}=\epsilon_{k_1k_2k_3}\epsilon_{def}\delta_{ijkk_1k_2}^{abcef}$. In terms of $\delta^\prime$, the fourth order contribution reads as $(\delta^\prime)_{ijkk_1}^{abcd}\ell_j^b\ell_k^c \ell_{k_1}^d$ which coincides with $\delta_{ijkk_1k_2}^{abcef}\ell_j^b\ell_k^c \ell_{k_1}^e\ell_{k_2}^f$ under the $D_\infty$ symmetry. $\delta^\prime$ thus plays the role of $\gamma$ and using the previous discussion about the third order contribution, we can then obtain its contribution to the dipole and octupole. Therefore, the purely spin-flipping part of the spin-orbit coupling contributes to the octupole at least at fourth order. This then completes the analysis of the leading order contribution to the octupole from different parts of the spin-orbit coupling.
	
	Armed with the above order-by-order analysis, we can now provide a general connection between Eq.~\eqref{eq_exp1} and Eq.~\eqref{eq_mexp}. The $D_\infty$ symmetry dictates that in the expansion in Eq.~\eqref{eq_exp1}, $\bm \ell^a$ with $a=1,2$ can only appear in two ways: $\bm \ell^1\times \bm \ell^2$ and $\sum_{a=1}^2 \ell_i^a\ell_j^a$. The first form is directly proportional to $\hat{\bm M}$. The second term is connected to $\hat{M}_i\hat{M}_j$ through the orthonormal condition of $\bm \ell^a$ as discussed previously. Combined with the fact that $\bm \ell^3=\hat{\bm M}$, Eq.~\eqref{eq_exp1} can always be transformed into a multipole expansion with respect to $\hat{\bm M}$. Furthermore, we note that terms that are at even order of $\hat{\bm M}$ necessarily contain an even times of appearance of $\bm \ell^3$ and $\bm \ell^1\times \bm \ell^2$, which is forbidden by the $D_\infty$ symmetry. This also shows the consistency of the $D_\infty$ symmetry with the Onsager's reciprocal relation. Therefore, Eq.~\eqref{eq_exp1} is fully equivalent to the multipole series in Eq.~\eqref{eq_mexp}.
	
	This equivalence then rigorously justifies that Eq.~\eqref{eq_mexp} is a multipolar expansion. Based on the above analysis, for an arbitrary pole with order $2a+1$, its leading order contribution has to come from the $2a+1$-th order expansion in Eq.~\eqref{eq_exp1}: it is the part of the $2a+1$-th order term that necessarily contains odd numbers of $\bm \ell^3$, possibly combined with $\sum_{a=1}^2 \ell_i^a\ell_j^a$ types of spin-flip part of the spin-orbit coupling. Purely spin-flip contribution starts from the $2a+2$-th order expansion. Therefore, the $2a+1$-th order pole is at least $2a+1$-th order in the spin-orbit coupling strength. This provides a solid ground for the multipole expansion in Eq.~\eqref{eq_mexp}.

      With our complete proof, it is now ready to discuss the difference between the multipolar expansion in Eq.~\eqref{eq_mexp} and previous empirical law in Eq.~\eqref{eq_emp}. An obvious difference is that the magnitude of $\bm M$ is now hidden in the various poles instead of acting as the expansion parameter, hence eliminating the divergence issue in the empirical law.  Moreover, our theory provides a clear picture of the emergence of the anomalous Hall effect: the poles can be expanded progressively with respect to the spin-orbit coupling strength; the role of the spin-flip and spin-converving part of the spin-orbit coupling in various poles can also be derived systematically. 
	
	We will end this section by discussing the contribution to the multipolar expansion due to the magnetic energy within our framework. We start with the expansion in Eq.~\eqref{eq_dE}.  The $D_{\infty}$ symmetry in the color space imposes the following constraints: first, same with the expansion of $\bm \sigma^H$, only three types of factors can appear, i.e., $\bm \ell^3$, $\bm \ell^1\times \bm \ell^2$, and $\ell_i^1\ell_j^1+\ell_i^2\ell_j^2$; Secondly, the power of $\bm \ell^3$ and $\bm \ell^1\times \bm \ell^2$ combined is even. The first constraint suggests that the expansion in Eq.~\eqref{eq_dE} is also a multipole expansion, and the second constraint eliminates any even-order terms. Therefore, Eq.~\eqref{eq_dE} can be put in the following form:
	\begin{align}
		E_M &= \tilde{B}_{ij} \hat{M}_i \hat{M}_j +\tilde{D}_{ijk\ell} \hat{M}_i \hat{M}_j \hat{M}_k \hat{M}_\ell+\cdots\,.
	\end{align}
	We note that the $2n$-th order term is at least of $2n$-order on the strength of the spin-orbit coupling.
	
	In general, the multipole in Eq.~\eqref{eq_mexp} depends on the energy $E$ and hence $E_M$. This dependence can be through the variation of the magnitude of $\bm M$ with the direction or the change of the energy or the chemical potential in the distribution function. Since $E_M$ is also a power series expansion over the spin-orbit coupling, its contribution to the anomalous Hall effect can be systematically included through expansion. For example, we consider the dipole contribution in Eq.~\eqref{eq_dip_sig}.  With the correction from $E_M$, we have
	\begin{align}\label{eq_em}
		\bm \sigma^{H,1}&=p_{ij}(E_0+E_M)\hat{M}_j \hat{e}_i\notag\\
		&=p_{ij}(E_0) \hat{M}_j \hat{e}_i+\frac{\partial p_{ij}}{\partial E_0} B_{k\ell} \hat{M}_j \hat{M}_k\hat{M}_\ell \hat{e}_i+\cdots\,.
	\end{align}
	The second term has the same form as the octupole contribution in Eq.~\eqref{eq_mexp} but the coefficient transform differently: the former has two rank-two tensors which transform independently while the latter transforms as a rank-four tensor as a whole. Other contributions from $E_M$ can be obtained straightforwardly using this recipe.

	\section{The dipolar structure}
	After establishing the full multipolar expansion theory of the anomalous Hall effect, we can now discuss its implication in the structure of the anomalous Hall effect. We first focus on the dipolar contribution which is linear in $\hat{\bm M}$, and it should replace the previous empirical law at the linear order. It implies two key information: first, the dependence of the anomalous Hall conductivity on the magnitude of the magnetization cannot generally be predicted, i.e., it is the direction of the magnetic order that directly matters; secondly, the structure of the anomalous Hall conductivity at the linear order is fully encoded by the dipole tensor. We want to further emphasize that despite the mutual linear dependence in our theory and in the empirical law, the former explicitly predict the dependence of the coefficient on the spin-orbit coupling, as illustrated later in Sect. III C.
	
	As discussed in the previous section, the leading order contribution to $\boldsymbol{\sigma}^H$ is from the dipole $p_{ij}$, which is at least of first order in the spin-orbit coupling. Since it is a rank-two tensor, it is convenient to decompose $p_{ij}$ into two parts, i.e., 
	\begin{align}
		\boldsymbol{p}=\boldsymbol{p}^s+\boldsymbol{p}^a,
	\end{align}
	where $\boldsymbol{p}^s$ and $\boldsymbol{p}^a$ are symmetric and antisymmetric part of $\boldsymbol{p}$, respectively. We refer to $\bm p_s$ as the symmetric dipole, and it is defined as $\boldsymbol{p}^s=(\boldsymbol{p}+\boldsymbol{p}^T)/2$. It can always be diagonalized by choosing suitable principal axes. We refer to $\boldsymbol{p}^a=(\boldsymbol{p}-\boldsymbol{p}^T)/2$ as the toroidal dipole, and it can not be diagonalized using orthogonal transformations. The independent elements of $\boldsymbol{p}^a$ transforms as a pseudovector labeled by $\bm d$ with its components given by
	\begin{align}
		d_i=\varepsilon_{ijk}p_{jk}\,.
	\end{align}
	
	The symmetric and toroidal dipole have distinct contributions to the anomalous Hall effect. In general, the anomalous Hall effect can have two different configurations as shown in Figs.~\ref{Fig_dipole_schematic}(a) and~\ref{Fig_dipole_schematic}(b).  For the conventional configuration in Fig.~\ref{Fig_dipole_schematic}(a), the magnetization is perpendicular to the Hall deflection plane showing a parallel relationship between $\bm M$ and $\boldsymbol{\sigma}^H$. This configuration is consistent with the empirical law in Eq.~\eqref{eq_emp}. In comparison, for the in-plane anomalous Hall effect in Fig.~\ref{Fig_dipole_schematic}(b), the magnetization is parallel to the Hall deflection plane corresponding to a vertical relationship between magnetization and $\boldsymbol{\sigma}^H$. These two types of anomalous Hall effect can coexist in materials and the resulting anomalous Hall conductivities can be decoupled through their distinct correlation with the magnetization direction. 
	The symmetric and toroidal dipole in the magnetic-order space provide the leading order source of anisotropy and hence an initial understanding of the requirements for these two configurations of the anomalous Hall effect, especially the in-plane one.
	
	\subsection{Symmetric dipole}

	\begin{figure}
		\includegraphics[width=8.5cm,angle=0]{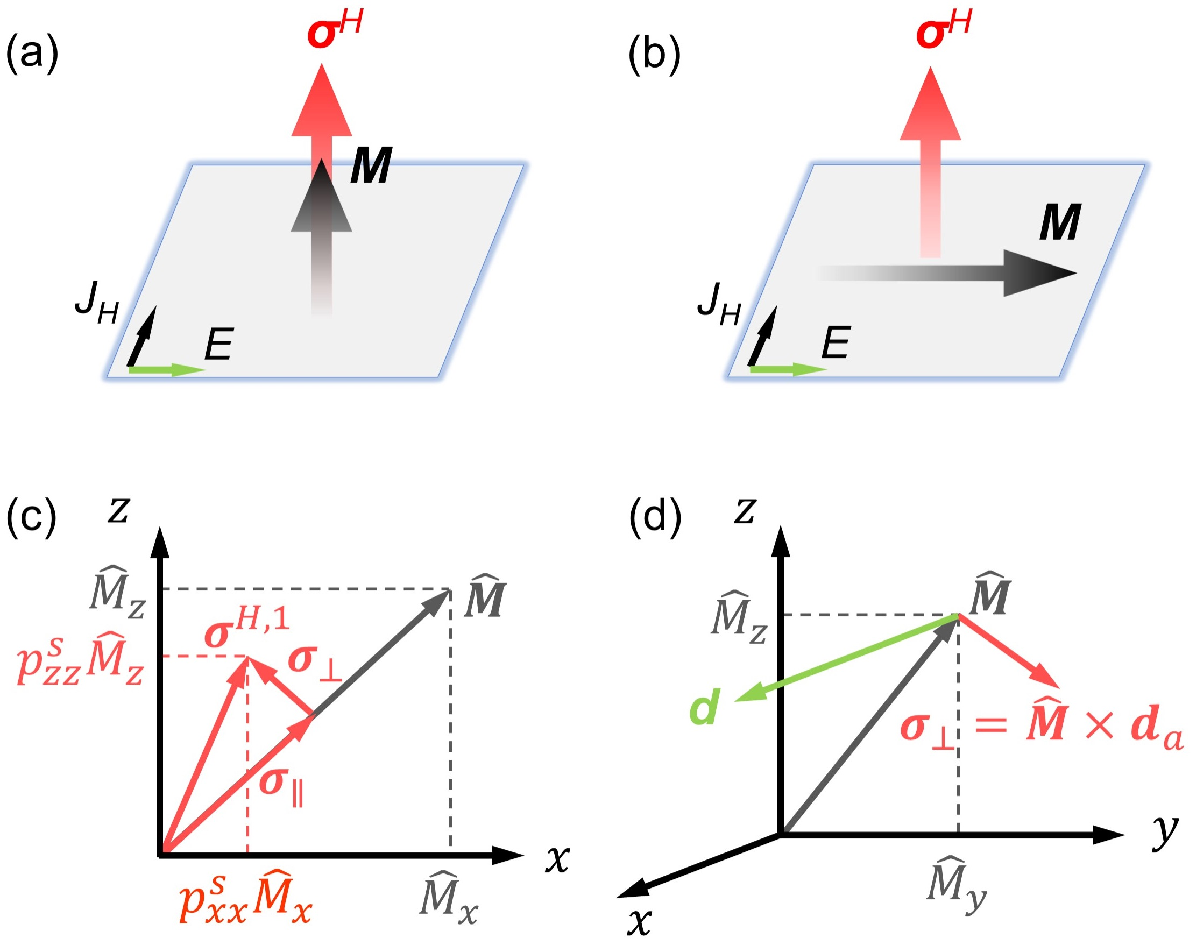}
		\caption{The two configurations of the anomalous Hall effect and their dipolar origin. (a) The conventional configuration with $\boldsymbol{\sigma}^H$ parallel to $\boldsymbol{M}$. (b) The in-plane anomalous Hall effect with $\boldsymbol{\sigma}^H$ perpendicular to $\boldsymbol{M}$. (c) the anomalous Hall effect from $\boldsymbol{p}^s$. $x$ and $z$ are two principal axes. (d) The anomalous Hall effect from $\boldsymbol{p}^a$. Only the in-plane component  $\boldsymbol{\sigma}_{\perp}$ exists and it is also perpendicular to $\boldsymbol{d}$.}
		\label{Fig_dipole_schematic}
	\end{figure}

	We start with the symmetric dipole $\bm p^s$. As a symmetric tensor, $\boldsymbol{p}^s$ can always be put in a diagonal form. In cubic systems with the point group symmetry given by  $T$, $T_d$, $T_h$, $O$, or $O_h$, since there are at least two different rotation axes with rotation angles smaller than $\pi$, $\bm p^s$ is isotropic and reduces to a scalar, i.e., $\boldsymbol{p}^s=p_0\boldsymbol{I}_{3\times3}$. In the other cases, the principal axes of $\bm p_s$ is usually given by the rotation axes or the normal direction of the mirror plane. Among them, there is a set of spacial cases where the crystal possesses a $C_n$ rotation symmetry with $n>2$. Then $\bm p^s$ has three nonzero elements  $p_{xx}^s$, $p_{yy}^s$, and $p_{zz}^s$, satisfying $p_{xx}^s=p_{yy}^s$. Therefore, $\bm p^s$ is isotropic in the plane normal to the rotation axes. We comment that since $\bm p^s$ is insensitive to the spatial inversion operation, $C_n$ is equivalent to $C_nI$ for $\bm p^s$. 
	
	With the symmetry constraint in mind, we can analyze its contribution to the anomalous Hall effect. In cubic system, $\bm p$ and hence $\bm p^s$ reduces to a scalar. As a result, $\boldsymbol{\sigma}^{H,1}$ is always parallel to $\boldsymbol{M}$, and the in-plane anomalous Hall effect does not exist. The previous empirical law always holds in this case. 
	
	In non-cubic system, the in-plane anomalous Hall effect emerges with the anisotropic $\bm p^s$. In this case, the three eigenvalues of $\bm p^s$ are different. For example, for crystals with a $C_{nz}$ axes~($n>2$), we generally have $p_{xx}^s=p_{yy}^s\neq p^s_{zz}$. Therefore, when $\hat{\bm M}$ is within $xy$ plane, $\bm p^s$ only brings the conventional configuration of the anomalous Hall effect. However, when $\hat{\bm M}$ is within $xz$ plane,  a misalignment between $\boldsymbol{\sigma}^{H,1}$ and $\boldsymbol{M}$ occurs due to the inequality of $p_{xx}^s$ and $p_{zz}^s$ as shown in Fig.~\ref{Fig_dipole_schematic}(c). The two configurations of the anomalous Hall effect then coexist, with $\bm \sigma_{\parallel}$ for the conventional one and $\bm \sigma_\perp$ for the in-plane one. Note that this type of anisotropy widely exist as long as $\boldsymbol{p}_s$ is anisotropic, as shown in  Table~\ref{table-1}.

    \begin{figure}
		\includegraphics[width=8.0cm,angle=0]{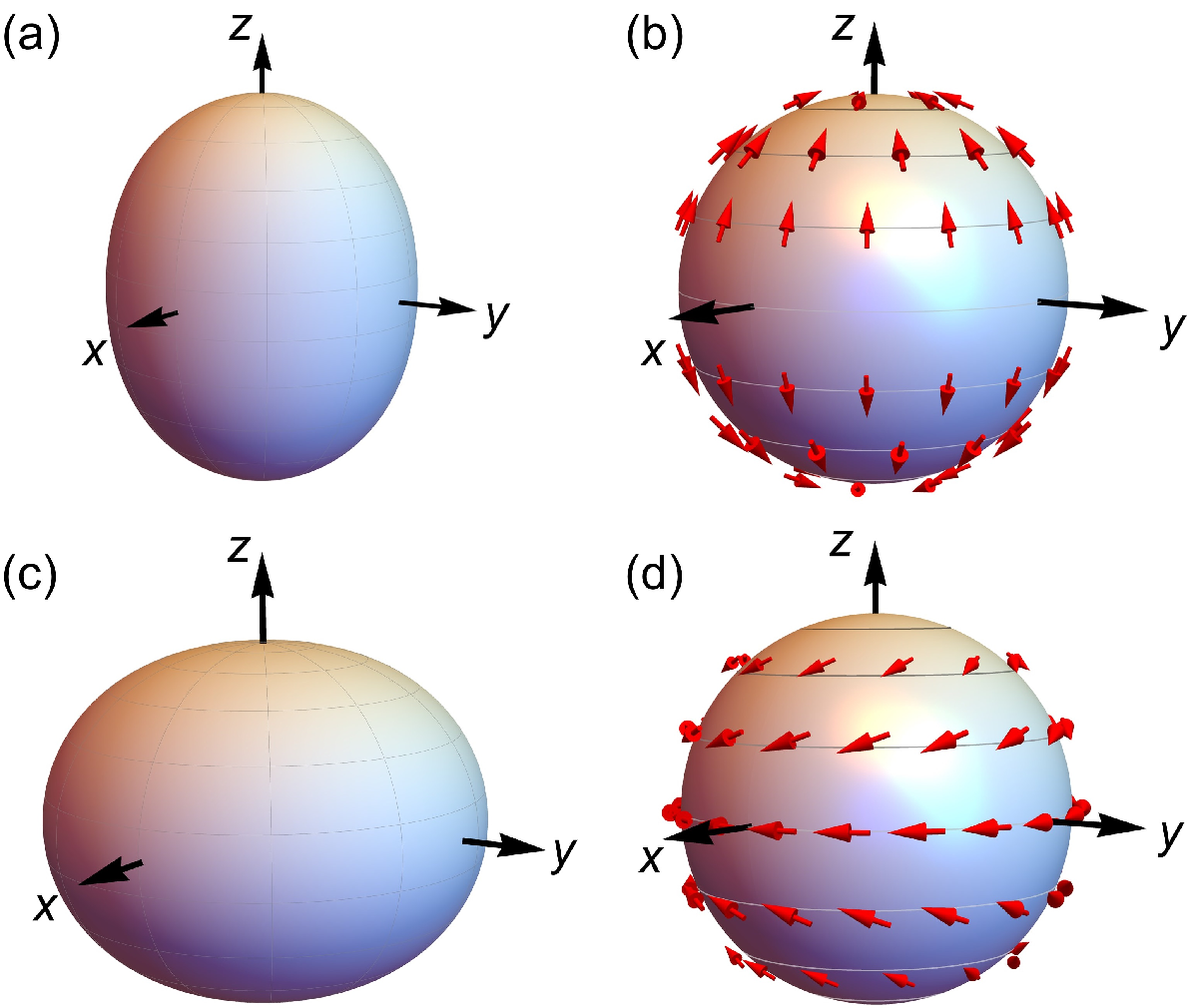}
		\caption{The dipolar structure of the anomalous Hall conductivity. Panel (a) and (b) show $\sigma_\parallel$ and $\bm \sigma_\perp$ in the $D_4$ group, respectively. Panel (c) and (d) show $\sigma_\parallel$ and $\bm \sigma_\perp$ in the $C_4$ group, respectively.  In (a) and (c), the azimuth angle and redial length respectively represents the direction and amplitude of $\sigma_{\parallel}$. In (b) and (d), arrows signify the direction and amplitude of $\boldsymbol{\sigma}_{\perp}$ while the azimuth angle indicates the magnetization direction.}
		\label{Fig_dipole_schematic2}
    \end{figure}

	The symmetric dipole endows the anomalous Hall conductivity with characteristic angular dependences when $\hat{\bm M}$ changes within certain planes. As a concrete example, we consider a crystal with $C_4$ point group symmetry. The nonzero elements of the symmetric dipole satisfies the following condition: $p_{xx}^s=p_{yy}^s\neq p_{zz}^s$ . When tilting the direction of magnetization from $z$ to $x$ by an angle of $\theta$, we expect the following profiles for the conventional part ($\sigma_{\parallel}$) and in-plane part ($\sigma_{\perp}$) of the anomalous Hall conductivity:    
	\begin{eqnarray}\label{eq:sigma_para_perp_dipole}
		\sigma_{\parallel}&=&\frac{p_{xx}^s+p_{zz}^s}{2}-\frac{p_{xx}^s-p_{zz}^s}{2}\cos 2\theta, \notag \\
		\sigma_{\perp}&=&-\frac{p_{xx}^s-p_{zz}^s}{2}\sin 2\theta.
	\end{eqnarray}
	The anisotropy in $\bm p^s$ affects the anomalous Hall conductivity in two ways: first, it makes $\sigma_{\parallel}$ angle-dependent; secondly, it leads to a nonzero $\sigma_\perp$. By fitting $\sigma_\parallel$ and $\sigma_\perp$ according to Eq.~\ref{eq:sigma_para_perp_dipole}, one can also get $p_{xx}^s$ and $p_{zz}^s$. 

To better illustrate the dipolar structure due to the symmetric dipole, we plot $\sigma_\parallel$ and $\bm \sigma_\perp$ as a function of $\hat{\bm M}$ for the $D_4$ group, as shown in Fig.~\ref{Fig_dipole_schematic2}(a) and~\ref{Fig_dipole_schematic2}(b), respectively. In the isotropic case, $\sigma_\parallel$ is a sphere and the dipolar structure can stretch or press the sphere into an ellipsoid. The pattern of $\bm \sigma_\perp$ clearly shows a dipolar structure. From Fig.~\ref{Fig_dipole_schematic2}(b), one can easily find the possible Hall-deflection plane that can show an in-plane anomalous Hall effect due to the symmetric dipole. The method is to draw a plane passing through the origin and focus on the intersecting great circle on the surface of the sphere. If there is any arrow that is not within the plane, the in-plane anomalous Hall effect is allowed. From this, one easily finds that to have the in-plane anomalous Hall effect, the Hall-deflection plane should not be perpendicular to any principal axis, as already discussed using symmetry analysis. 
	
	\subsection{Toroidal dipole}
	We now analyze the symmetry constraint on the toroidal dipole. Since $\bm d$ is an axial vector, any rotation axes will pin the $\bm d$ vector to the same direction and any mirror plane will pin $\bm d$ to the normal direction of the plane. Therefore, $\bm p_a$ is zero in cubic systems. It can only exist in crystals with at most one rotation axis, or one mirror plane. If both rotation axis and mirror plane exist, they should be normal to each other. There are ten point groups supporting $\bm p^a$, as shown in  Table~\ref{table-1}.

	Interestingly, the toroidal dipole can only induces the in-plane anomalous Hall effect, as illustrated in Fig.~\ref{Fig_dipole_schematic}(d). Based on the definition, we have
	\begin{align}
		\boldsymbol{\sigma}^{H,1}=\hat{\boldsymbol{M}}\times\boldsymbol{d}\,.
	\end{align}
	Therefore, the Hall deflection plane is spanned by $\hat{\bm M}$ and $\bm d$. We still take the crystal with $C_4$ symmetry as an example. The toroidal dipole satisfies: $p_{xy}^a=-p_{yx}^a=d_z$. When $\hat{\bm M}$ varies in the $xz$ plane, the toroidal dipole contributes to the in-plane anomalous Hall conductivity in the following way: 
	\begin{align}
		\bm \sigma_\perp=-\hat{e}_y d_z \sin\theta\,.
	\end{align}
	When $\hat{\bm M}$ changes in the $xy$ plane, since $\bm p^s$ is isotropic, we have
	\begin{align}
		\sigma_{\parallel}=p_{xx}^s\,, \sigma_{\perp}=d_z\,.
	\end{align}
	Interestingly, both $\sigma_{\parallel}$ and $\sigma_\perp$ are angle-independent.
	
	We are now in a position to discuss the whole dipolar contribution to the anomalous Hall effect. In experiment, the Hall plane is usually fixed with $\hat{\bm M}$ easily tunable by a large magnetic field. We will focus on the more exotic in-plane setup where $\hat{\bm M}$ varies within the Hall deflection plane. From the properties of $\bm p$, we find that to make the symmetric dipole contribute, the Hall deflection plane should not be perpendicular to any principal axes and to make the toroidal dipole contribute, the Hall deflection plane should not be perpendicular to the $\bm d$ vector. Take the crystal with $C_4$ point group as an example. For $\sigma_z^H$, there is no in-plane contribution. For $\sigma_x^H$, when $\hat{\bm M}$ changes from $\hat{y}$ to $\hat{z}$, we have
	\begin{align}
		\sigma_x^H=d_z\cos\theta\,.
	\end{align}
	In a more general case where the Hall deflection plane makes a $\pi/4$ angle relative to both the $xy$ and $xz$ plane, when the magnetization makes a $\theta$ angle with $x$ axis in the Hall deflection plane, we have
	\begin{eqnarray}\label{eq:sigma_para_perp_dipole_exp}
		\sigma_{\perp}=-\frac{p_{xx}^s-p_{zz}^s}{2}\sin\theta+\frac{\sqrt{2}}{2}d_z\cos\theta.
	\end{eqnarray}
	One immediately finds that both $\bm p^s$ and $\bm p^a$ contributes in general and they have distinct angle dependences.

To visualize a full dipolar structure, we plot $\sigma_\parallel$ and $\bm \sigma_\perp$ as a function of $\hat{\bm M}$ for the $C_4$ group, as shown in Fig.~\ref{Fig_dipole_schematic2}(c) and~\ref{Fig_dipole_schematic2}(d), respectively. Both the symmetric part and the toroidal part is nonzero. From Fig.~\ref{Fig_dipole_schematic2}(c), we find that the toroidal dipole does not affect $\sigma_\parallel$, which is still in the form of an ellipsoid. In sharp contrast, Fig.~\ref{Fig_dipole_schematic2}(d) demonstrates that in $\bm \sigma_\perp$, besides the clear dipolar structure from the symmetric part, the toroidal dipole introduces an additional chiral structure, which is most easily seen in the equator, where the arrow form a head-to-tail alignment. This shows the toroidal nature. One can still judge if a particular Hall-deflection plane can support the in-plane setup by studying the arrow on the intersecting great circle between the Hall-deflection plane and the surface of the sphere. But it is straightforward to find out that only the $xy$ plane cannot in this case. 
	
	In summary, the dipole in the magnetization space offers the leading order interpretation of the configuration of the anomalous Hall effect. Its elements can be probed from the characteristic angular dependence of $\bm \sigma^H$. Specifically, we show how to obtain the in-plane anomalous Hall effect from the anisotropic symmetric part and antisymmetric part of the dipole. Our theory is consistent with recent experimental and theoretical progress in the study of the in-plane anomalous Hall effect\cite{Cao2023,Zhou2022,Tan2021}. In strained CuMnAs\cite{Cao2023}, VS$_2$-VS superlattice\cite{Cao2023,Zhou2022}, and ferrimagnetic Weyl semimetal FeCr$_2$Te$_4$\cite{Tan2021}, the lattices only have one generalized two-fold rotation symmetry along $y$ axis, i.e., $C_{2y}$ or $M_y$, thus the dipole $\boldsymbol{p}$ will take the following form
    \begin{align}
		\bm p=\begin{pmatrix}
			p^s_{xx} & 0 &p_{xz}^s-d_y\\
			0 & p^s_{yy} &0 \\
			p_{xz}^s+d_y & 0 & p^s_{zz}
		\end{pmatrix}\,.
    \end{align}
    Based on this dipolar structure, we find that the in-plane anomalous Hall effect in these materials comes from the off-diagonal element $p_{xz}^s\pm d_y$. Besides offering a unified picture, our theory also greatly enriches the material candidate and provides experimental guidance for the in-plane anomalous Hall effect.
	
	\subsection{Lattice model}

	To illustrate the dipolar structure, we utilize a lattice with the $D_{4h}$ point group. The tight-binding model is constructed using the three $t_{2g}$ orbitals $d_{xy}$, $d_{yz}$, and $d_{zx}$. Up to the next-nearest neighbor hopping, the tight-binding Hamiltonian for a single spin species can be expressed as\cite{supp}
	\begin{eqnarray}\label{eq:t2g1}
		H_0=
		\begin{pmatrix}
			A_{xy}   &   B_{xz}   & B_{yz} \\
			B_{xz}   &   A_{yz}   & B_{xy} \\
			B_{yz}   &   B_{xy}   & A_{zx}
		\end{pmatrix},
	\end{eqnarray}
	where $A_{ij}=-2t(\cos k_i+\cos k_j+s\delta_{jz}\cos k_j)$ and $B_{ij}=4t'\sin k_i\sin k_j$ with $t$ and $t'$ being nearest and next-nearest neighbors hopping strength, respectively. $s$ is an anisotropy parameter accounting for the difference between the $z$ direction and the $x$ and $y$ directions. Taking into account the spin-orbit coupling and exchange coupling, we can write the full Hamiltonian as
	\begin{eqnarray}\label{eq:t2g2}
		H=
		\begin{pmatrix}
			H_0+\frac{\lambda}{2} L_z &   \frac{\lambda}{2}(L_x-iL_y) \\
			\frac{\lambda}{2}(L_x+iL_y)   &  H_0-\frac{\lambda}{2} L_z
		\end{pmatrix}+J\hat{\bm m}\cdot \bm \sigma,
	\end{eqnarray}
	where $\lambda$ is the strength of spin-orbit coupling, $J$ and $\hat{\boldsymbol{m}}$ are the coupling strength and direction of magnetization, respectively. The angular momentum operator $\boldsymbol{L}$ for the $t_{2g}$ orbitals are given by
	\begin{eqnarray}\label{eq:Lx}
		L_x=
		\begin{pmatrix}
			0 &   0   & -i \\
			0 &   0   &  0 \\
			i &   0   &  0
		\end{pmatrix},
	\end{eqnarray}
	\begin{eqnarray}\label{eq:Ly}
		L_y=
		\begin{pmatrix}
			0 &   i   &  0 \\
			-i &   0   &  0 \\
			0 &   0   &  0
		\end{pmatrix},
	\end{eqnarray}
	\begin{eqnarray}\label{eq:Lz}
		L_z=
		\begin{pmatrix}
			0 &   0   &  0 \\
			0 &   0   &  i \\
			0 &  -i   &  0
		\end{pmatrix}.
	\end{eqnarray}
	
	\begin{figure}
		\includegraphics[width=8.5cm,angle=0]{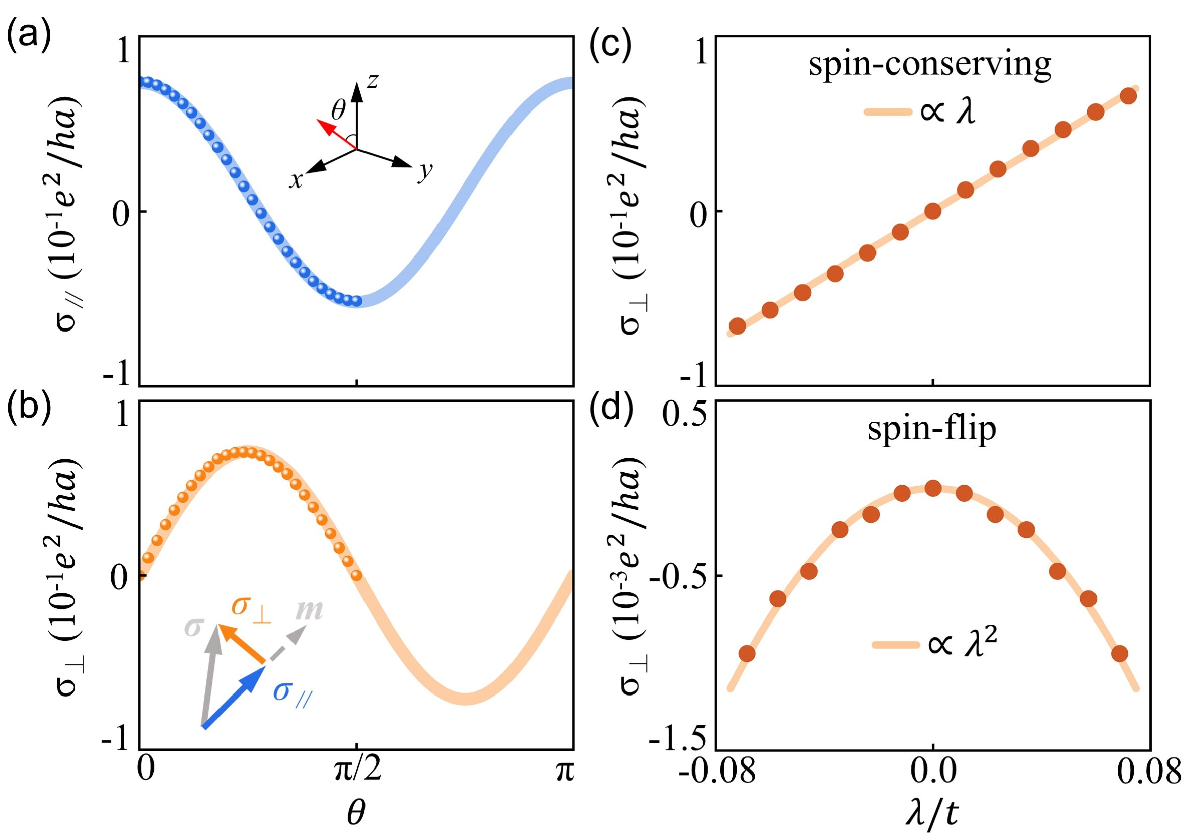}
		\caption{The dipolar structure in a tight-binding model with $D_{4h}$ point group symmetry. The parameters of this model are $s=0.2$, $t'/t=0.1$, $\lambda/t=0.08$, and $J/t=0.5$. The lattice constant is labeled by $a$. (a) and (b) separately show the conventional and in-plane part of anomalous Hall conductivity with $\hat{\boldsymbol{m}}$ making a $\theta$ angle with $\hat{z}$ in the $xz$ plane. The dots are the direct calculation and the solid curves show the fitting to Eq.~\ref{eq:sigma_para_perp_dipole}. The insets illustrate the definition of the $\sigma_{\parallel}$ and $\sigma_{\perp}$. The Fermi level is $\mu=0$. (c) and (d) show the dependence of the in-plane anomalous Hall conductivity at $\theta=\pi/4$ on different parts of the spin-orbit coupling.  In (d), the Fermi-level is shifted to $\mu/t=3.2$ to obtain a relatively large $\sigma_{\perp}$.}
		\label{Fig_dipole_model}
	\end{figure}
	
	More generally, for arbitrary magnetization direction, we use the previous definition of the spin-orbit vectors, and express the spin-orbit coupling term as follows
	\begin{eqnarray}\label{eq:soc}
		\hat{H}_{soc}=\lambda\sum_{a}(\boldsymbol{\ell}^a\cdot \boldsymbol{L})\sigma_a,
	\end{eqnarray}
	where $\boldsymbol{\ell}^a$ is spin-orbit vector shown in Eq.~\ref{eq_socvec2}. This form helps the separation of the spin-conserving and spin-flip part of the spin-orbit coupling: the term with $\boldsymbol{\ell}^3$ is the spin-conserving part and those with $\boldsymbol{\ell}^1$ and $\boldsymbol{\ell}^2$ are the spin-flip part.
	
	We then investigate the configuration of the anomalous Hall effect (Calculation methods are shown in the Supplemental Material~\cite{supp}). The conventional and in-plane part is shown in Figs.~\ref{Fig_dipole_model}(a) and~\ref{Fig_dipole_model}(b), respective. On the other hand, with $D_{4h}$ point group symmetry, the dipole $\boldsymbol{p}$ only has symmetric part, satisfying the condition $p^s_{xx}=p^s_{yy}\neq p^s_{zz}$. When $\hat{\bm m}$ changes from $\hat{z}$ to $\hat{x}$ by an angle of $\theta$, we expect that $\sigma_{\parallel}$ and $\sigma_{\perp}$ obey Eq.~\ref{eq:sigma_para_perp_dipole}.
	By fitting the dots to Eq.~\ref{eq:sigma_para_perp_dipole}, we find that in unit of $10^{-3}e^2/ha$,  $(p^s_{xx}-p^s_{zz})/2=-62.7$ from $\sigma_{\parallel}$, and $(p^s_{xx}-p^s_{zz})/2=-71.0$ from $\sigma_{\perp}$. These results are quite close, demonstrating the validity of our theory.
	
	A major feature of our theory is that we can predict the exact dependence of the dipole on different types of the spin-orbit coupling. To validate our theory, we fixed the magnetization direction at $\theta=\pi/4$. In Fig.~\ref{Fig_dipole_model}(c), we plot the $\sigma_{\perp}$ versus $\lambda$ with $\hat{H}_{soc}=\lambda(\boldsymbol{\ell}^3\cdot \boldsymbol{L})\sigma_3$, i.e., when only the spin-conserving part of spin-orbit coupling is left. It is evident that $\sigma_{\perp}$ scales linearly with $\lambda$. In Fig.~\ref{Fig_dipole_model}(d), the spin-orbit coupling takes the form $\hat{H}_{soc}=\lambda\sum_{a}(\boldsymbol{\ell}^a\cdot\boldsymbol{L})\sigma_a$ with $a=1,~2$, i.e., only the spin-flip part is left. It can be found that $\sigma_{\perp}$ scales quadratically with $\lambda$ in this case. These scaling properties are consistent with our theory that the spin-conversing part contributes to the dipole at first order and the spin flip part contributes at second order.

	\subsection{Dipolar structure in MnGa}

	\begin{figure}[t]
		\includegraphics[width=8.5cm,angle=0]{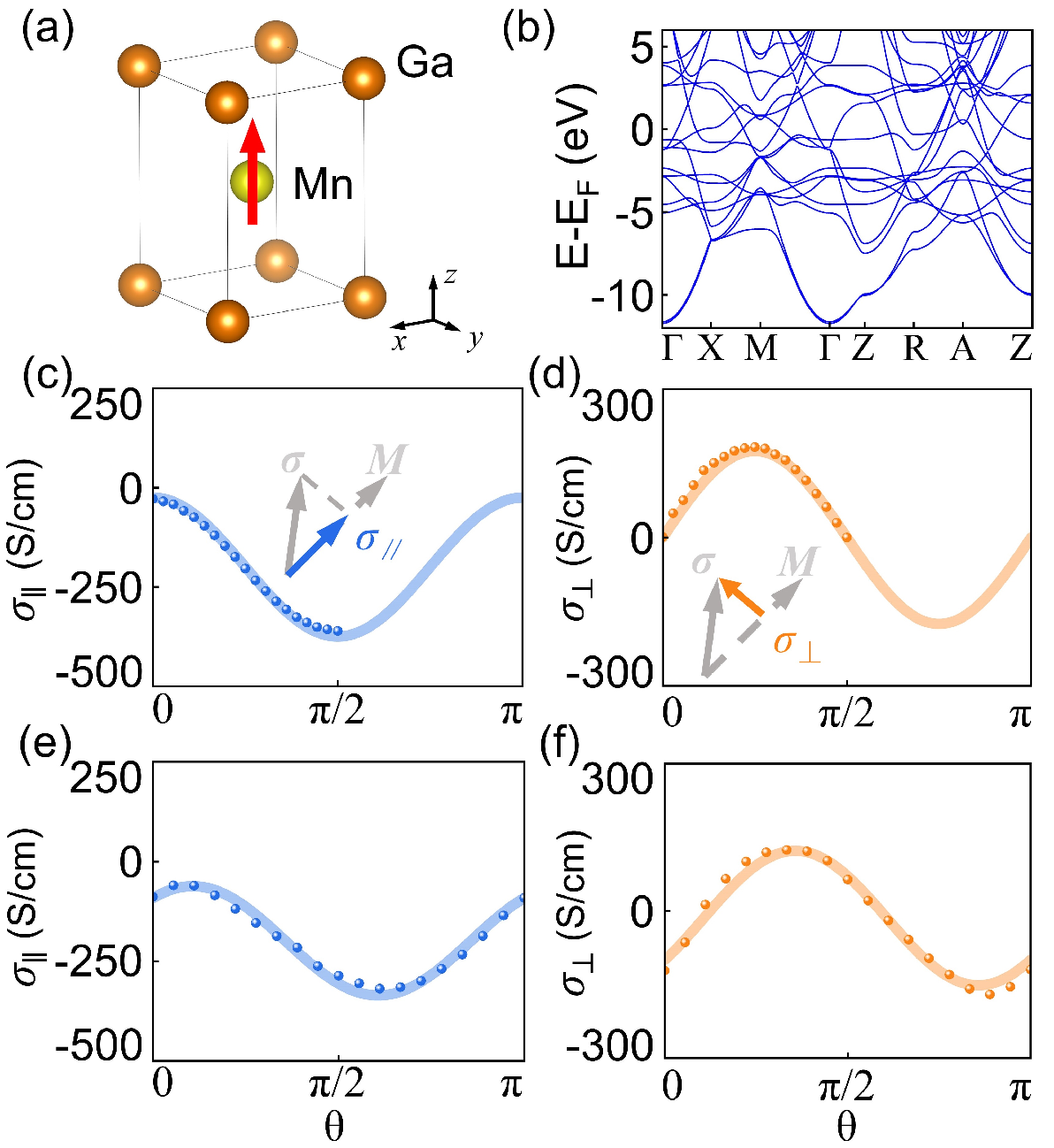}
		\caption{The anomalous Hall conductivity in MnGa. (a) The unit cell of MnGa. (b) The band structure. (c) and (d) The $\sigma_{\parallel}$ and $\sigma_{\perp}$ in MnGa with $\boldsymbol{M}$ being tilted from $z$ to $x$ by an angle of $\theta$. (e) and (f) The $\sigma_{\parallel}$ and $\sigma_{\perp}$ in strained MnGa, with the same angle defined as in (c) and (d). $E_F=-0.2$~eV is used for the calculation.} \label{Fig_MG}
	\end{figure}
	
	To show the predictive power of our theory, we demonstrate the dipolar structure in real materials. 
	We consider $\delta$-MnGa, a ferromagnetic metal whose electronic properties and grown method have been systematically studied\cite{XUESHAN1980,Tanaka1993,Yang1998,Ryan2017}. As shown in Fig.~\ref{Fig_MG}(a), MnGa has a body centered tetragonal crystal structure which belongs to $D_{4h}$ point group.  As discussed before, the nonzero elements of the dipole should satisfy  $p^s_{xx}=p^s_{yy}\neq p^s_{zz}$. Therefore, the in-plane anomalous Hall effect is expected from the anisotropy in $\bm p$. The band structure of MnGa with magnetization along $z$ direction is shown in Fig.~\ref{Fig_MG}(b). Its band structure clearly exhibits metallic characteristics, where the primary contributions around the Fermi level stem from Mn $d$ orbitals and Ga $p$ orbitals. (Calculation methods are shown in the Supplemental Material~\cite{supp}).
	
	To show the dipolar structure, one need to rotate the direction of magnetization. Our calculation shows that the magnetic anisotropy is under 0.2~meV throughout the magnetization direction rotation, suggesting an easy control of direction via an external magnetic field. The $\sigma_{\parallel}$ and $\sigma_{\perp}$ after rotation of magnetization direction  in the $xz$ plane are shown in dots in Fig.~\ref{Fig_MG}(c) and ~\ref{Fig_MG}(d), respectively. The first-principles data fits Eq.~\ref{eq:sigma_para_perp_dipole} very well as can be seen from the solid curves. From $\sigma_\parallel$, we have  $(p^s_{xx}-p^s_{zz})/2=-164.6$ for $\sigma_{\parallel}$; from $\sigma_\perp$ we have $(p^s_{xx}-p^s_{zz})/2=-186.6$. They are quite close, verifying again our theory in real materials. We note that the in-plane anomalous Hall effect is quite large in this case: $\sigma_\perp$ reaches a value of 186.6~S/cm at $\theta=\pi/4$, which is comparable to $\sigma_{\parallel}$. 
	
	
    Originally, the antisymmetric part of $\boldsymbol{p}$ vanishes as there are two different rotation axes in $D_{4h}$. We then apply a shear strain in the $x$ direction to reduce the symmetry of MnGa to $C_{2h}$.  The dipole $\bm p$ then takes the following form
	\begin{align}
		\bm p=\begin{pmatrix}
			p^s_{xx} & 0 &p_{xz}^s-d_y\\
			0 & p^s_{yy} &0 \\
			p_{xz}^s+d_y & 0 & p^s_{zz}
		\end{pmatrix}\,.
	\end{align}
	Therefore, both symmetric and antisymmetric part of $\boldsymbol{p}$ will contribute to $\sigma_\perp$ in this case.
	When $\hat{\bm M}$ makes a $\theta$ angle with the $z$ axis in the $xz$ plane, we derive the following angular dependence of $\sigma_{\parallel}$ and $\sigma_{\perp}$:
	\begin{eqnarray}\label{eq:sigma_para_perp_dipole2}
		\sigma_{\parallel}&=&\frac{p^s_{xx}+p^s_{zz}}{2}-b\cos (2\theta+\theta_0), \notag\\
		\sigma_{\perp}&=&d_y-b\sin (2\theta+\theta_0).
	\end{eqnarray}
	Here $\theta_0$ is due to the misalignment of the principal axis with the $x$ and $z$ axes with $b\sin(\theta_0) = p_{xz}^s$ and $b\cos(\theta_0) = (p^s_{xx} - p^s_{zz})/2$. Comparing with Eq.~\ref{eq:sigma_para_perp_dipole},  $\sigma_{\perp}$ involves a constant term from $d_y$, which signifies the unique contribution of the antisymmetric part of the dipole.  Evidently, $d_y$ is not involved in $\sigma_{\parallel}$, in accordance with Fig.~\ref{Fig_dipole_schematic}(d). From Figs.~\ref{Fig_MG}(e) and~\ref{Fig_MG}(f),  the first-principles results~(dots) fit Eq.~\eqref{eq:sigma_para_perp_dipole2}~(solid curve) very well. From $\sigma_\parallel$ we have $(p^s_{xx}+p^s_{zz})/2=-198.3$~S/cm, $b=-126.3$~S/cm, $\theta_0=5.60$. From $\sigma_\perp$, we have $d_y=-14.5$~S/cm, $b=-148.1$~S/cm, $\theta_0=5.63$. The value of $b$ and $\theta$ is quite close from these two different sets of data, showing the validity of our theory.

	\section{The octupolar structure}
      Although the dipolar structure constitutes the leading order contribution to the anomalous Hall effect, the octupolar structure introduces the leading order nonlinearity of $\bm \sigma^H$ in the magnetic-order space. The key object, i.e., the octupole $o_{ijk\ell}$, transforms as a rank-four tensor and is hence subject to the following constraint:
    \begin{eqnarray}\label{eq:octupole_constraint}
		o_{ijkl}=S_{ii^\prime}S_{jj^\prime}S_{kk^\prime}S_{\ell \ell^\prime}o_{i^\prime j^\prime k^\prime \ell^\prime},
    \end{eqnarray}
where $S_{ii^\prime}$ is the matrix representation of the symmetry operation. Generally speaking, as the rank of a tensor increases, it is more sensitive to various symmetries and hence supports much richer structures. Specifically, the octupole $o_{ijk\ell}$ has the following striking properties.

\begin{figure}
		\includegraphics[width=8.5cm,angle=0]{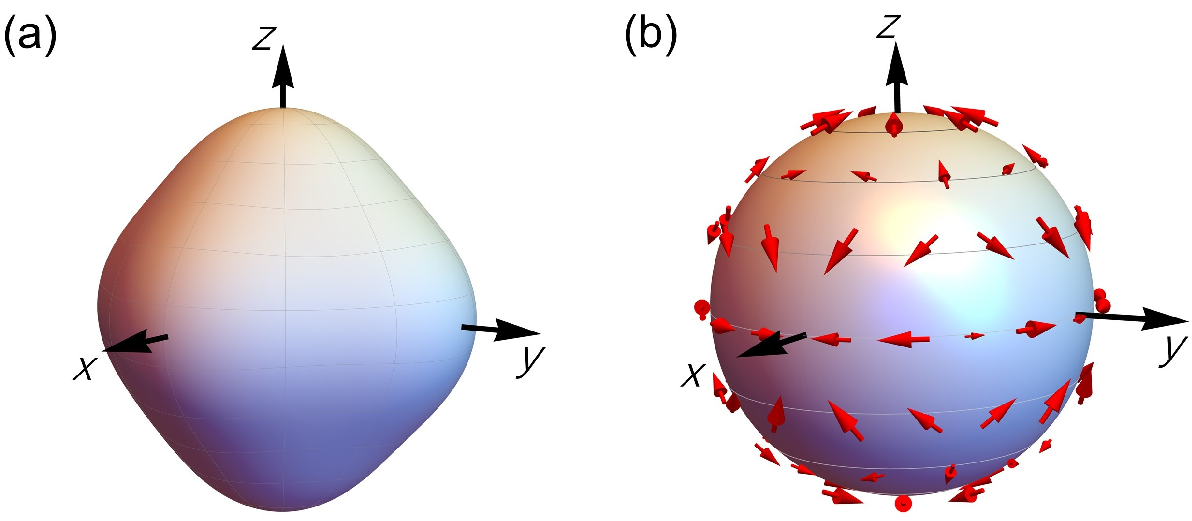}
		\caption{The octupolar structure of the anomalous Hall conductivity in $O_h$ point group. (a) and (b) separately illustrate $\sigma_{\parallel}$ and $\bm \sigma_{\perp}$. In (a), the azimuth angle and redial length respectively represents the direction and amplitude of $\sigma_{\parallel}$. In (b), arrows signify the direction and amplitude of $\boldsymbol{\sigma}_{\perp}$ while the azimuth angle indicates the magnetization direction.} \label{Fig_octupole_schematic}
\end{figure}

First, although there are fewer independent elements with higher point-group symmetry, the octupole is always allowed to have nonzero elements in principle. As a result, nonlinearity always exists in the anomalous Hall conductivity.

Secondly, as discussed previously, since $o_{ijk\ell}$ is traceless with respect to the last three indices, its contribution to $\bm \sigma^H$ is always different from that from the dipole. When the magnetization direction is allowed to vary, this difference should be reflected by different angle-dependent profiles. This is thus the first method to differentiate the octupolar structure from the dipolar one both theoretically~(numerically) and experimentally. We emphasize that in this case, the dipole and octupole coexists and the former one dominates.

Thirdly, the in-plane anomalous Hall effect is inherent in the octupolar structure due to the nonlinearity.  In fact, there exists unique octupolar anisotropy when the dipolar structure does not lead to the in-plane anomalous Hall effect but the octupolar structure does. The inclusion of the octupole thus greatly expands the possibility of the in-plane anomalous Hall effect. This is the second method to differentiate the octupolar structure from the dipolar one, and will be the focus of our following discussions.

	\subsection{Unique octupolar anisotropy: cubic }

	\begin{figure}
		\includegraphics[width=8.5cm,angle=0]{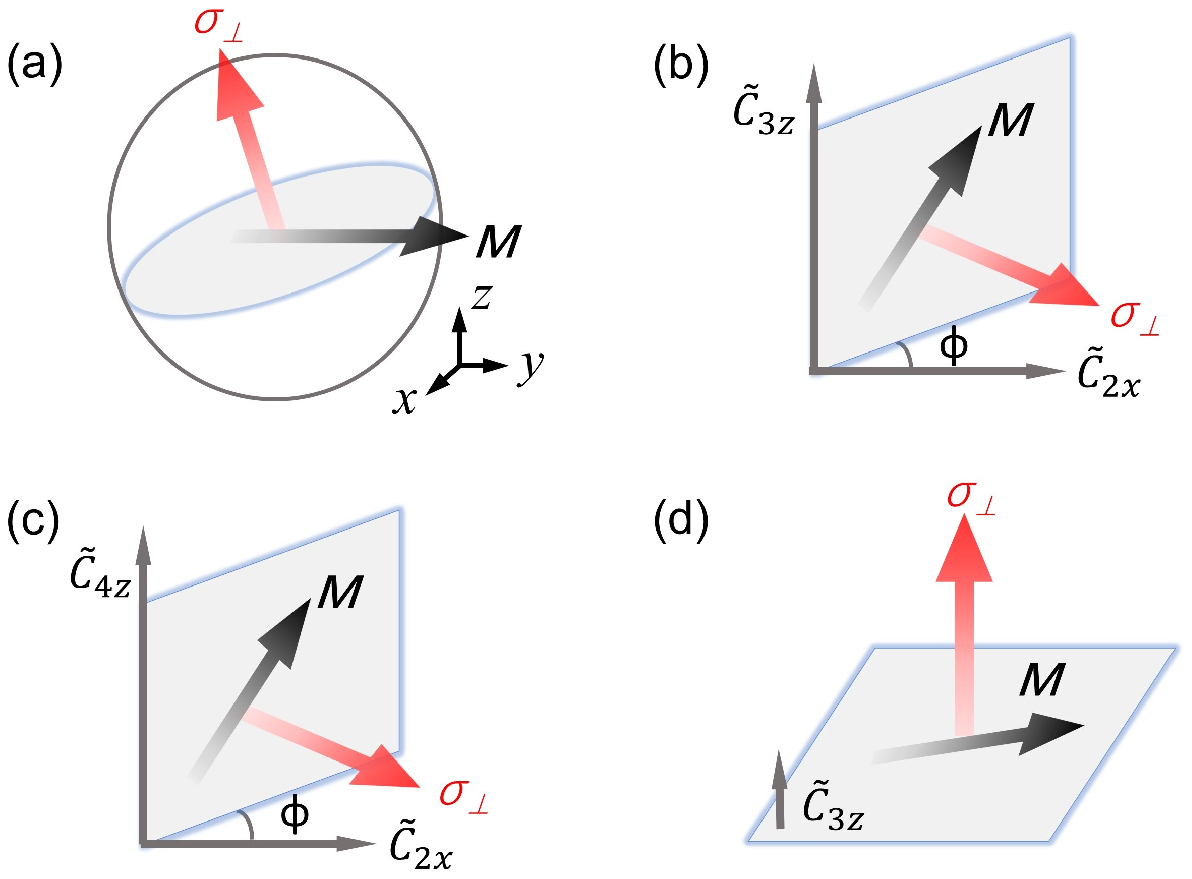}
		\caption{The four configurations for in-plane anomalous Hall effect purely from octupole. (a) The type $O$-I configuration in cubic lattice. The Hall plane is arbitrary in cubic lattice. (b) The type $O$-II structure with three equivalent generalized rotation axes which are connected by $\tilde{C}_{3z}$. The Hall plane is perpendicular to the $xy$-plane. (c) The type $O$-III structure with two equivalent generalized rotation axes which are connected by $\tilde{C}_{4z}$. The Hall plane is also perpendicular to the $xy$-plane. (d) The type $O$-IV configuration, which requires that the lattice exhibit $\tilde{C}_{3z}$ but lack $\tilde{C}_{2z}$ symmetry.} \label{Fig_octupole_schematic2}
	\end{figure}

    To discuss the unique in-plane anomalous Hall effect due to the octupolar structure, we rely on Table.~\ref{table-1} and focus on the cases where the dipolar structure cannot lead to the in-plane configuration. We start from the five cubic groups, which contains at least two rotation axis with $n>2$. We refer to the octupolar anisotropy in the cubic group as the $O$-I type~(see Table~\ref{table-2}). In this case, the dipole reduces to a scalar and its contribution to the anomalous Hall conductivity always follow $\hat{\bm M}$. In sharp contrast, the nonlinearity from the octupole still brings anisotropy. For the $T_d$, $O$, and $O_h$ group, a generic rank-4 tensor has four independent elements. Since the octupole $o_{ijk\ell}$ is symmetric with respect to the last three indices, the number of independent elements reduces to $2$. The traceless condition further reduces this number to $1$. We note that this is the smallest possible number of independent elements among all point groups.  The nonzero elements satisfy the following condition: $o_{xxxx}=o_{yyyy}=o_{zzzz}=-2o_{xxyy}=-2o_{xxzz}=-2o_{yyxx}=-2o_{yyzz}=-2o_{zzxx}=-2o_{zzyy}$. 
    
    To see the octupolar structure, we follow the previous treatment and decompose $\bm \sigma^H$ into $\sigma_\parallel$ and $\bm \sigma_\perp$. In Fig.~\ref{Fig_octupole_schematic}, we illustrate $\sigma_\parallel$ and $\bm \sigma_\perp$ as a function of $\hat{\bm M}$ for a generic choice of strength of the dipole and octupole element. For $\sigma_\parallel$, the dipole contribution alone would make the plot a perfect sphere. The deviation from a sphere thus clearly demonstrates the octupolar structure. More importantly, $\bm \sigma_\perp$ vanishes with the dipole contribution alone, while the octupolar structure introduces a nontrivial pattern based on the point group symmetry.  It is straightforward to find out that along any high-symmetry direction, $\bm \sigma_\perp$ vanishes. The pattern thus shows the transition from one high-symmetry direction to another. 
    
    More precisely, the anomalous Hall conductivity under $T_d$, $O$ and $O_h$ group can be put in the following form
    \begin{eqnarray}\label{eq:sigma_Oh}
		\sigma^H_i=\alpha \hat{M}_i+\beta \hat{M}_i^3,
    \end{eqnarray}
    with $\alpha=p_0-\frac{3}{2}o_{xxxx}$ and $\beta=\frac{5}{2}o_{xxxx}$.  Therefore, the anomalous Hall conductivity only depends on the magnetization along the same direction, although the dependence is nonlinear. We align $\hat{\bm M}$ with one high-symmetry direction, e.g., (001) direction, and then rotate it to another high-symmetry direction, e.g., the $(100)$ direction. The angle dependence of $\sigma_{\parallel}$ and $\sigma_{\perp}$ read as
    \begin{eqnarray}\label{eq:O_sigma_para_perp}
    \sigma_{\parallel}&=&p_0+\frac{3}{8}o_{xxxx}+\frac{5}{8}o_{xxxx}\cos4\theta, \notag \\
    \sigma_{\perp}&=&\frac{5}{8}o_{xxxx}\sin4\theta.
    \end{eqnarray}
    Compared with the dipolar structure, the transition due to the octupole is in the form of fourth harmonics. 
    
   We now find the guiding rule for the experimental observation of the in-plane anomalous Hall effect in materials with $T_h$, $O$ and $O_h$ point group.  According to  Fig.~\ref{Fig_octupole_schematic}(b), we can draw the interested Hall-deflection plane passing through the origin and find its intersecting great circle on the sphere surface. If the arrow has any perpendicular component, the in-plane anomalous Hall effect is allowed in this setup. According to previous discussion, as long as the normal direction of the plane deviates from the high-symmetry directions, the in-plane anomalous Hall effect appears, as illustrated in Fig.~\ref{Fig_octupole_schematic2}(a). 
    
     As a concrete example, we take the (103) plane as the Hall deflection plane. We assume $\hat{\bm M}$ varies in this plane, and label the angle between it and (010) as $\theta$. The anomalous Hall conductivity then reads as
     \begin{align}
     \sigma^H=\frac{3}{5}o_{xxxx}\sin^3\theta\,.
     \end{align}
     We note that this type of angle dependence has already been observed in experiments in bcc iron\cite{Peng2024}.

    As discussed in Sec. II D, the magnetic energy also contributes to the anomalous Hall conductivity. Specifically, based on Eq.~\eqref{eq_em}, the dipole structure combined with the quadrupole term in the magnetic energy is cubic in $\hat{M}$ and hence seems to have the same form as the octupole. However, since $p_{ij}$ and $B_{k\ell}$ transform independently, one immediately finds that this dipole-quadrupole mixed structure is quite different from the octupole contribution. Under $T_d$, $O$ and $O_h$ group, both $p_{ij}$ and $B_{k\ell}$ reduce to a scalar. Therefore, this mixed structure is identical to the dipole contribution, whose contribution to $\sigma_\parallel$ is isotropic and whose contribution to $\sigma_\perp$ vanishes identically.
    
    Finally, we shall mention that all the above discussion can be generalized to the other two cubic groups, i.e., $T$ and $T_h$ group. In this case, $o_{ijk\ell}$ has two independent elements, i.e., $o_{xxxx}$ and $o_{xxyy}$. The anomalous Hall conductivity then reads as
    \begin{align}
    \sigma_i^H=\alpha^\prime \hat{M}_i +\beta^\prime \hat{M}_i^3+\gamma^\prime \hat{M}_i \hat{M}_{i+1}^2\,,
    \end{align}
    where $\alpha^\prime=p_0-3(o_{xxxx}+o_{xxyy})$, $\beta^\prime=4o_{xxxx}+3o_{xxyy}$, $\gamma^\prime=3(o_{xxxx}+2o_{xxyy})$, and $(i,i+1)$ takes the value $(x,y)$, $(y,z)$, and $(z,x)$. In the case that $o_{xxxx}=-2o_{xxyy}$ as in the $T_d$, $O$, and $O_h$ group, $\alpha^\prime$ and $\beta^\prime$ reduce to $\alpha$ and $\beta$, respectively, and $\gamma^\prime=0$.
    The octupolar anisotropy can then be derived based on this set of equations. The guiding rule for experimental observation is similar and the magnetic anisotropy energy still does not mix with the octupolar structure.

\subsection{Unique octupolar anisotropy: single rotation }

	\begin{table}[t]
		\caption{Four classes of point groups with which only the octupole can contribute to certain types of the in-plane anomalous Hall effect.}
		\begin{ruledtabular}
			\begin{tabular}{ccc}
				type           &  key symmetry elements                                  & point group  \\ \hline    
		         $O$-I        & $\tilde{C}_{nx}$ and $\tilde{C}_{nx^\prime}~(n>2)$ &  $T$, $T_d$, $T_h$, $O$, $O_h$ \\
				$O$-II         & $\tilde{C}_{3z}$ and $\tilde{C}_{2x}$      & $C_{3v}$, $D_3$, $D_{3d}$ \\
				$O$-III        & $\tilde{C}_{4z}$ and $\tilde{C}_{2x}$      &   $C_{4v}$, $D_4$, $D_{4h}$ \\ 
                $O$-IV          & $\tilde{C}_{3z}$                                     & $C_3$, $S_6$, $C_{3v}$, $D_3$, $D_{3d}$ \\
         
			\end{tabular}
		\end{ruledtabular}
		\label{table-2}
	\end{table}

All the other cases where the in-plane anomalous Hall effect solely comes from the octupolar structure involves a single rotation axis with $n>2$. Since the inversion symmetry does not affect the multipole, the operation $C_n$ and $C_n I$ are equivalent, and we shall label this generalized rotation by $\tilde{C}_n$.

We shall start from the case where there is an additional two-fold rotation $\tilde{C}_2$ perpendicular to the principal rotation axis. This include the group $C_{nv}$, $D_n$, and $D_{nh}$ with $n=3,4,6$ and $D_{3d}$. In this case,  the toroidal dipole vanishes, and only the symmetric dipole contributes. The principal axis for $p_{ij}^s$ are the $z$ axis and any direction in the $xy$ plane. As discussed previously, if the Hall-deflection plane contains any two different principal axis, the in-plane anomalous Hall effect vanishes. We first focus on the plane parallel to the principal rotation axis and we find that the octupolar structure can make the in-plane anomalous Hall effect emerge but the result is highly sensitive to the degree of rotation. 

We first consider the $C_{3v}$, $D_3$ and $D_{3d}$ group, which has a $\tilde{C}_{3z}$ and a $\tilde{C}_{2x}$ symmetry. The nonzero elements of the octupole satisfies the following conditions: $o_{xxxx}=o_{yyyy}=3o_{xxyy}=3o_{yxxy}=-3/4o_{xxzz}=-3/4o_{yyzz}$, $o_{xxyz}=o_{yxxz}=-o_{yyyz}$, $o_{zzzz}=-2o_{zxxz}=-2o_{zyyz}$, and $o_{zyyy}=-o_{zxxy}$. If the Hall-deflection plane makes an angle $\phi$ with the $x$ axis, and $\hat{\bm M}$ rotates within this plane, forming an angle $\theta$ with the $z$ axis, as shown in Fig.~\ref{Fig_octupole_schematic2}(b), we then have
    \begin{eqnarray}\label{eq:O_II_sigma_perp}
    \sigma_{\perp}&=&3o_{xxyz}\cos3\phi\sin^2\theta\cos\theta.
    \end{eqnarray}
The in-plane anomalous Hall is solely induced by octupole element $o_{xxyz}$ and it is required that the Hall-deflection plane is not perpendicular to any $\tilde{C}_2$ axis. We refer to this type of anisotropy as the $O$-II case~(see Table~\ref{table-2}). It is straightforward to check that the cubic contribution from the dipole-quadrupole mixed structure from the magnetic anisotropy energy does not contribute in this case. This is due to the fact that $p_{xz}=p_{yz}=0$. 

We proceed to consider the $C_{4v}$, $D_4$ and $D_{4h}$ group, which has a $\tilde{C}_{4z}$ and a $\tilde{C}_{2x}$ symmetry. The nonzero elements of the octupole satisfies the following conditions: $o_{xxxx}=o_{yyyy}=-o_{xxyy}-o_{xxzz}=-o_{yxxy}-o_{yyzz}$, $o_{xxyy}=o_{yxxy}$, $o_{xxzz}=o_{yyzz}$, and $o_{zzzz}=-o_{zxxz}-o_{zyyz}$.  As shown in Fig.~\ref{Fig_octupole_schematic2}(c), with $\phi$ and $\theta$ defined in the same way as in the O-II case, we have
    \begin{eqnarray}\label{eq:O_III_sigma_perp}
    \sigma_{\perp}&=&\frac{o_{xxxx}-3o_{xxyy}}{4}\sin4\phi\sin^3\theta.
    \end{eqnarray}
It is still required that the Hall-deflection plane is not perpendicular to any $\tilde{C}_2$ axis to allow the in-plane anomalous Hall effect. We refer to this type of anisotropy as the $O$-III case~(see Table~\ref{table-2}). The magnetic anisotropy energy still does not contribute in this case.

It is worth comparing the anisotropy in the $O$-II with that in the $O$-III case. Symmetry wise, the only difference is that the degree of rotation is $3$ in the $O$-II case but $4$ in the $O$-III case. However, this single difference enforces entirely different constraints on the octupole element: in the $O$-II case, $o_{xxxx}=3o_{xxyy}$, while in the $O$-III case, $o_{xxyz}=0$. Therefore, although a unique in-plane anomalous Hall effect is allowed with similar experimental setup, the angle dependence is quite different. This difference can be further demonstrated by allowing the $\hat{\bm M}$ change in the $xy$ plane, and decomposing the in-plane component of $\bm \sigma^H$ into $\sigma_\parallel$ and $\sigma_\perp$. We then have
    \begin{eqnarray}\label{eq:O_sigma_para_perp2}
    \sigma_{\parallel}&=&p_{xx}+\frac{3(o_{xxxx}+o_{xxyy})}{4}+\frac{o_{xxxx}-3o_{xxyy}}{4}\cos4\theta, \notag \\
    \sigma_{\perp}&=&\frac{o_{xxxx}-3o_{xxyy}}{4}\sin4\theta.
    \end{eqnarray}
We then find that for groups with $\tilde{C}_{3z}$, $\bm \sigma^H$ is still isotropic within the $xy$ plane even with the octupolar structure, but groups with $\tilde{C}_{4z}$ is not.

For the group $C_{6v}$, $D_6$, $D_{6h}$ with a $\tilde{C}_{6z}$ and $\tilde{C}_{2x}$, $o_{xxyz}=0$ and $o_{xxxx}=3o_{xxyy}$. Therefore, both $O$-II and $O$-III type of anisotropy does not exist. The in-plane anomalous Hall effect as shown in Fig.~\ref{Fig_dipole_schematic}(b) does not exist. 

We now consider the case where the Hall-deflection plane is perpendicular to the principal rotation axis, i.e., the $xy$ plane, as shown in Fig.~\ref{Fig_octupole_schematic2}(d). The dipolar structure does not contribute to this type of the in-plane anomalous Hall effect since $p_{zx}=p_{zy}=0$. For groups with $\tilde{C}_{4z}$ and $\tilde{C}_{6z}$, a $\tilde{C}_{2z}$ is implied and the in-plane anomalous Hall effect with the setup shown in Fig.~\ref{Fig_octupole_schematic2}(d) will always vanish. This is because $\tilde{C}_{2z}$ will reverse the sign of the magnetization and keep $\sigma_z^H$ unchanged, which violates the Onsager relation that $\boldsymbol{\sigma}^H$ is an odd function of magnetization. The remaining point group then includes $C_3$, $S_6$, $C_{3v}$, $D_3$, and $D_{3d}$. The relevant octupole elements satisfy: $o_{zxxx}=-o_{zxyy}$ and $o_{zyyy}=-o_{zxxy}$. We comment that for the $C_{3v}$, $D_3$ and $D_{3d}$ group, there is an additional constraint that $o_{zxxx}=0$. The anomalous Hall conductivity along the $z$ direction then reads as
    \begin{eqnarray}\label{eq:O_sigma_perp}
    \sigma_{\perp}&=&o_{zxxx}\cos3\theta-o_{zyyy}\sin3\theta.
    \end{eqnarray}
Therefore, the octupolar structure can indeed induce this type of the in-plane anomalous Hall effect and we will refer to it as the $O$-IV case~(see Table~\ref{table-2}). It is simple to check that the magnetic anisotropy energy does not contribute in this case.

	\begin{figure}
		\includegraphics[width=8.5cm,angle=0]{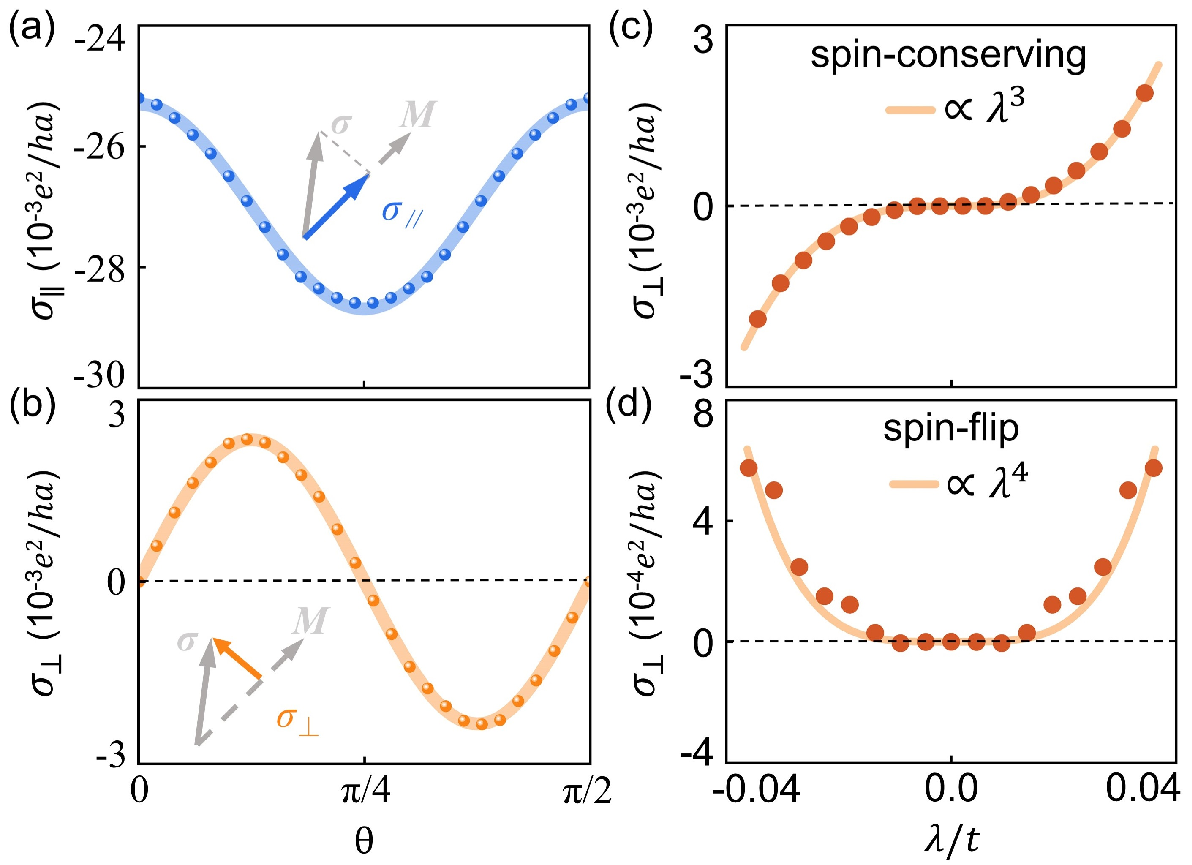}
		\caption{The octupolar structure for a tight-binding cubic lattice model. (a) and (b) show $\sigma_{\parallel}$ and $\sigma_{\perp}$ with $\boldsymbol{M}$ being tilted from $z$ to $x$ by an angle of $\theta$, respectively. $\sigma_{\parallel}$ and $\sigma_{\perp}$ are illustrated in the insets. The Hamiltonian is given by Eq.~\ref{eq:t2g2}, with parameters set as follows: $t'/t=0.1$, $s=0$, $\lambda/t=0.04$, and $|\boldsymbol{M}|/t=0.5$. (c) and (d) show $\sigma_{\perp}$ at $\theta=\pi/8$ as a function of the spin-conserving part and spin-flip part of spin-orbit coupling, respectively. The dots show the calculated data and the solid line represents the fitted function. In (d), the Fermi-level is shifted from electrically neutral point $E/t=0.0$ to $E/t=3.1$ to obtain a relatively large value.}
		\label{Fig_octupole_model}
	\end{figure}

    \subsection{Lattice model}

    As a concrete example, we consider a tight-binding model in a cubic lattice. The Hamiltonian is given by Eq.~\ref{eq:t2g2} with the anisotropy parameter $s=0$. The point group is thus $O_h$. As discussed previously, the dipole structure is isotropic, but the octupolar structure is not. Specifically, the leading order contribution to the in-plane anomalous Hall effect comes from the latter.

    Figures~\ref{Fig_octupole_model}(a) and~\ref{Fig_octupole_model}(b) show the angle dependence of $\sigma_{\parallel}$ and $\sigma_{\perp}$ with $\hat{\bm m}$ changing from $\hat{z}$ to $\hat{x}$. Both of them have a period of $\pi/2$, consistent with our symmetry analysis in Sec. IV A. They also fit well to Eq.~\ref{eq:O_sigma_para_perp} (solid curve) with $p_0=-28.01$, $o_{xxxx}=2.71$ for $\boldsymbol{\sigma}_{\parallel}$ and $o_{xxxx}=3.75$ for $\boldsymbol{\sigma}_{\perp}$, in unit of $10^{-3}e^2/ha$. The octupole element $o_{xxxx}$ is indeed about one order of magnitude smaller than $p_0$, but it determines the strength of anisotropy, highlighting its importance in the characterization of the structure of anomalous Hall effect. 
    
    To further check the validity of our theory, we study the scaling of the octupole with the strength of the spin-orbit coupling. We fix the magnetization direction so that $\hat{\bm m}$ makes a $\pi/8$ angle with the $\hat{z}$ axis. Similar with the treatment in Sec.III C, we decompose the spin-orbit coupling into the spin-conserving part and the spin-flip part. Since the in-plane anomalous Hall effect is proportional to $o_{xxxx}$, we then plot $\sigma_\perp$ as a function of the spin-orbit coupling strength, with only the spin-conserving part in Fig.~\ref{Fig_octupole_model}(c) and spin-flip part in Fig.~\ref{Fig_octupole_model}(d). Evidently, $\sigma_{\perp}$ scales with $\lambda^3$ for spin-conserving part and $\lambda^4$ for spin-flip part of spin-orbit coupling. Together with the first and second order scaling behavior in Figs.~\ref{Fig_dipole_model}(c) and~\ref{Fig_dipole_model}(d), this provides a strong support of our theory.

    \begin{figure}
		\includegraphics[width=8.5cm,angle=0]{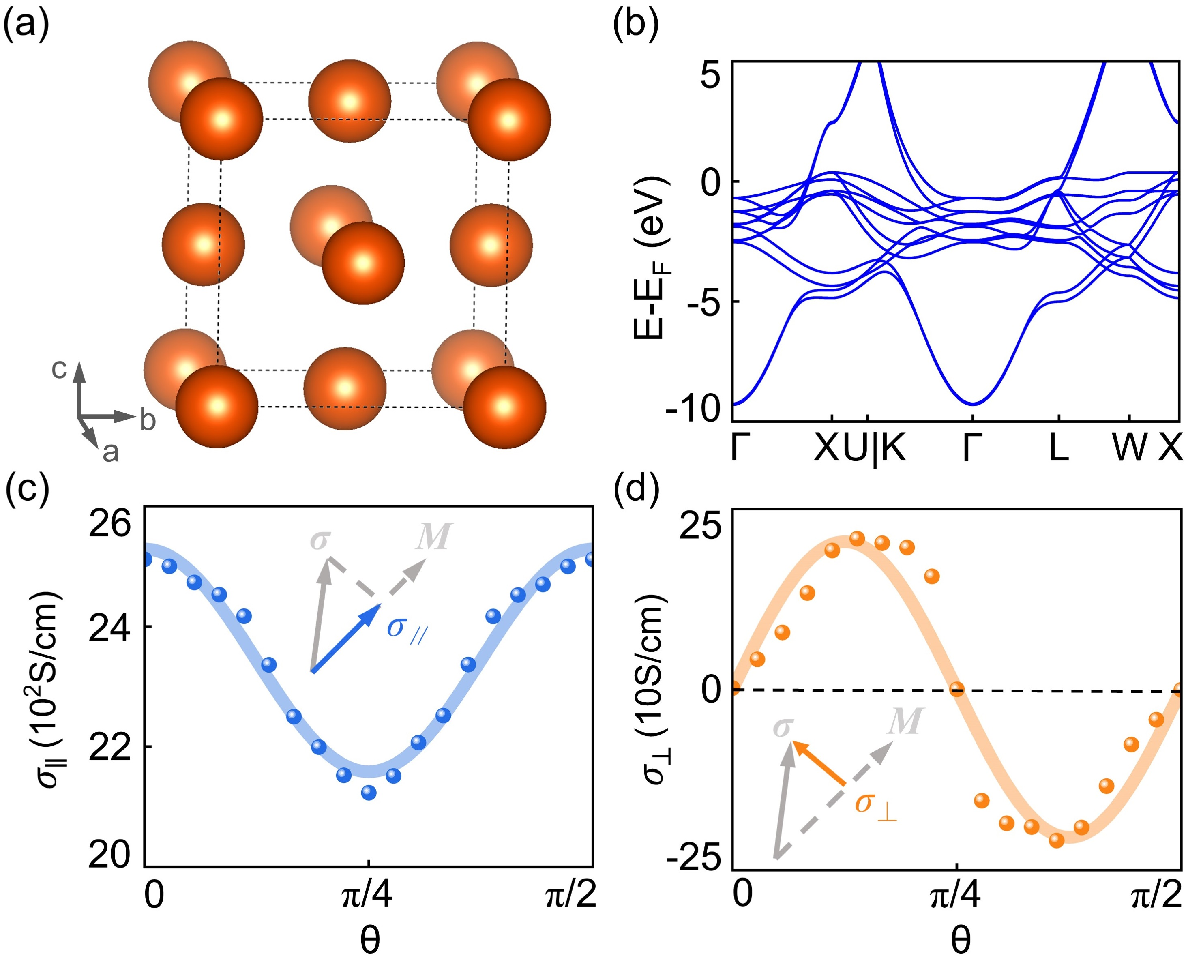}
		\caption{The lattice, band structure, and anomalous Hall conductivity of fcc Ni.  (a) The lattice structure of fcc Ni. (b) The band structure of fcc Ni for the magnetization along $z$ direction. (c) and (d) The parallel part $\sigma_{\parallel}$ and in-plane part $\sigma_{\perp}$ of anomalous Hall conductivity with magnetization is tilted from $z$ to $x$ by an angle of $\theta$. The dots are data from first-principles calculations and the solid lines are fittings according to Eq.~\ref{eq:O_sigma_para_perp}.}
		\label{Fig_Fe}
    \end{figure}

    \subsection{Cubic octupolar structure in fcc Ni}

    \begin{figure*}
	\includegraphics[width=14.5cm,angle=0]{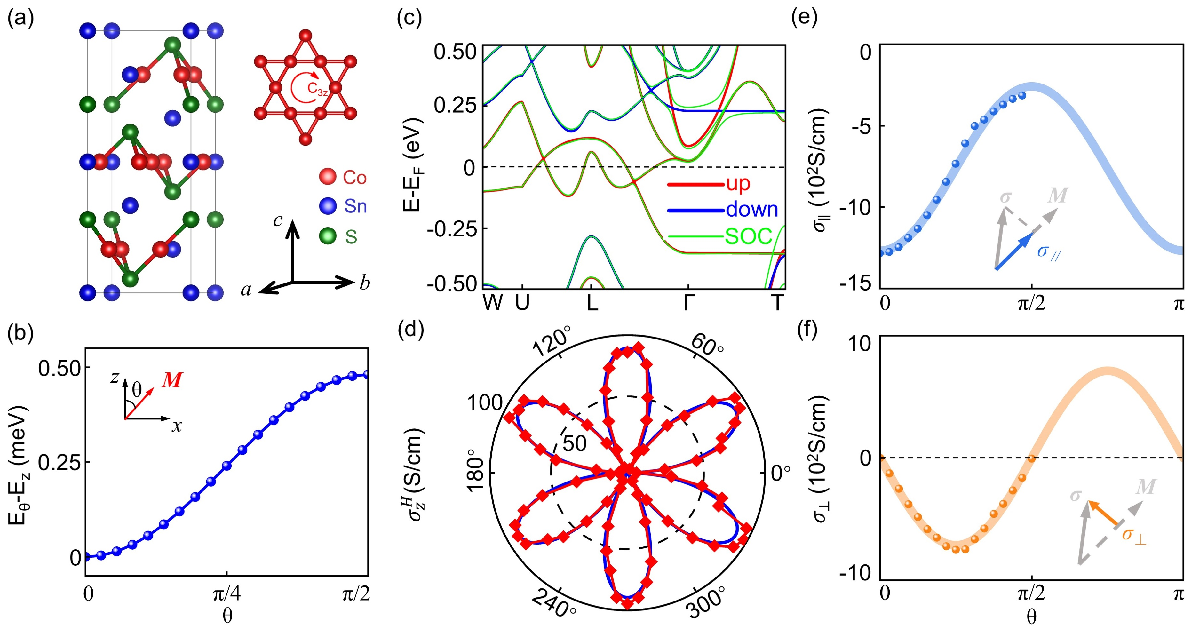}
	\caption{The crystal, magnetic properties, and anomalous Hall conductivity multipolar structure of Co$_3$Sn$_2$S$_2$. (a) The lattice structure of Co$_3$Sn$_2$S$_2$. The Cobalt atoms form a Kagome lattice with $C_{3z}$ rotation symmetry. (b) The magnetic anisotropic energy of Co$_3$Sn$_2$S$_2$. $E_\theta$ represents the total energy for the magnetization at an angle $\theta$, whereas $E_z$ signifies the total energy when magnetization aligns with the $z$-axis. (c) The band structure for the magnetization along $x$ direction. The bands of spin up, spin down, and spin-orbit coupling are marked as up, down, and SOC, respectively. (d) The anomalous Hall conductivity $\sigma^H_z$ as a function of magnetization direction $\varphi$. The magnetization is constrained in $xy$-plane and tilted counterclockwise from $x$ axis. The red dots and line represent the first-principles data, while the blue curve depicts the fitted results. (e) and (f) separately plot the $\sigma_{\parallel}$ and $\sigma_{\perp}$ with $\boldsymbol{M}$ is tilted from $c$ to $b$ by an angle of $\theta$. The solid lines are fits to the first-principles data.}
	\label{Fig_CSS}
    \end{figure*}

   We now demonstrate the octupolar structure of $\bm \sigma^H$ in real materials using first-principles calculations. We first focus on the $O$-I type anisotropy in cubic groups. The material we choose is fcc Ni with a Curie temperature $T_c\approx600$~K\cite{Zhang2001}. Although it has been studied previously~\cite{Meyer2010,Lalmi2012}, the octupolar structure and the possibility of the in-plane anomalous Hall effect have not been thoroughly explored. Figure~\ref{Fig_Fe}(a) shows the lattice of Ni which belongs to $O_h$ point group. The band structure with spin-orbit coupling for the magnetization along $z$ axis is shown in Fig.~\ref{Fig_Fe}(b). Around the Fermi-level, the band is mainly composed of $d$ orbitals. 
	
    To show the octupolar structure, we rotate $\hat{\bm M}$ in the $xz$-plane, and calculate $\sigma_{\parallel}$ and $\sigma_{\perp}$ as a function of the angle $\theta$ are shown in Figs.~\ref{Fig_Fe}(c) and~\ref{Fig_Fe}(d), respectively. (Calculation methods are shown in the Supplemental Material~\cite{supp}). Both exhibit a period of $\pi/2$, in sharp contrast with the $\pi$ period observed in MnGa (see Figs.~\ref{Fig_MG}(c) and~\ref{Fig_MG}(d)). This difference illustrates distinct multipolar structures in these two materials. 

    We can further extract the dipole and octupole element from these two profiles in fcc Ni. As shown in Figs.~\ref{Fig_Fe}(c) and~\ref{Fig_Fe}(d), the first-principles data fits well to Eq.~\ref{eq:O_sigma_para_perp} with the fitting parameters given by $p_0=2232.4$~S/cm, $o_{xxxx}=296.2$~S/cm for $\sigma_{\parallel}$ and $o_{xxxx}=327.3$~S/cm for $\sigma_{\perp}$. The octupole element is order-of-magnitude smaller than the dipole one as expected. However, the in-plane anomalous Hall conductivity originating from the octupole can still reach around 200 S/cm which is sufficient from experimental detection.

    \subsection{Octupolar structure with single rotation in Co$_3$Sn$_2$S$_2$}

We now demonstrate the octupolar anisotropy with a single rotation. The material we choose is the ferromagnetic Weyl semimetal Co$_3$Sn$_2$S$_2$, which hosts both a large anomalous Hall conductivity and a giant anomalous Hall angle\cite{Liu2018_2,Wang2018}. The methods for single crystal growth and transport measurement of this material have been well established in experiments\cite{Liu2018_2,Wang2018,XiaoLin2012,Schnelle2013,Kassem2015}. The lattice structure of Co$_3$Sn$_2$S$_2$ is shown in Fig.~\ref{Fig_CSS}(a). It is crystallized in a hexagonal lattice with point group of $D_{3d}$ which possesses one $C_{3}$ rotation symmetry along $c$ axis and three equivalent $C_2$ axes in $ab$-plane. For convenience, we use Cartesian coordinates to label the directions, with $x$, $y$, and $z$ axis corresponding to the $b_{\perp}$, $b$, and $c$ direction, respectively. In this notation, one of the $C_2$ axis is along $x$ direction.

   The magnetic anisotropic energy is plotted in Fig.~\ref{Fig_CSS}(b). One can see that the $z$ axis is the easy axis of magnetization and the magnetic anisotropy energy between $z$ axis and $x$ axis is $\Delta=E_z-E_x=-0.41$~meV. Such a tiny magnetic anisotropy indicates that the magnetization can be rotated into $xy$-plane by using an external magnetic field. The band structure with the magnetization along $x$ axis is shown in Fig.~\ref{Fig_CSS}(c). In the absence of spin-orbit coupling, one can see that only one spin channel crosses the Fermi-level and the other one is gapped. After introducing the spin-orbit coupling, band gaps are opened for same-spin bands and opposite-spin bands, revealing the role of the spin-conserving and spin-flip part of spin-orbit coupling.
       
        The $D_{3d}$ point group symmetry enforces that the dipole of $\bm \sigma^H$ in Co$_3$Sn$_2$S$_2$ only has the symmetric part with $p^s_{xx}=p^s_{yy}\neq p^s_{zz}$. This leads to the anisotropy when the magnetization rotates in the $xz$ plane. Using first-principle calculations, we plot $\sigma_{\parallel}$ and $\sigma_{\perp}$ in this situation, as shown in Figs.~\ref{Fig_CSS}(d) and~\ref{Fig_CSS}(e). The data fits well with Eq.~\ref{eq:sigma_para_perp_dipole}. (Calculation methods are shown in the Supplemental Material~\cite{supp}).
        
    We now discuss the octupolar structure. According to Table.~\ref{table-2}, the octupolar structure in $D_{3d}$ group can exhibit the $O$-II and $O$-IV type of anisotropy. We choose the Hall-deflection plane to be the $xy$ plane to demonstrate the $O$-IV type of anisotropy. The anomalous Hall conductivity $\sigma^H_z$ should have the following profile when the magnetization rotates in the $xy$ plane:
    \begin{eqnarray}\label{eq:sigma_c3}
    	\sigma^H_{z}=-O_{zyyy}\sin3\theta.
    \end{eqnarray}
    The data from first-principles calculations fit well with this relation, from which we can extract that $O_{zyyy}=81.6$~S/cm, clearly demonstrating the octupole structure in Co$_3$Sn$_2$S$_2$.

     Recently, an in-plane anomalous Hall effect with a period of $2\pi/3$ was observed in Fe$_3$Sn$_2$ by rotating the magnetization direction in the Hall-deflection plane~(i.e., the $xy$-plane)\cite{Wang2024}. Since the point group of Fe$_3$Sn$_2$ is $D_{3d}$, this periodicity clearly demonstrates that the origin of this in-plane anomalous Hall effect  is from the octupole element $o_{zyyy}$.


	

	\section{Discussion}
	
	In summary, we have rigorously established the relation between the anomalous Hall conductivity and the ferromagnetic order, which  is in the form of a multipolar expansion. Based on our theory, the previous empirical law in Eq.~\eqref{eq_emp} should be replaced by the following relation
	\begin{align}
	\sigma_i^H=\sigma_0 H_i+p_{ij}\hat{M}_j+o_{ijk\ell} \hat{M}_j\hat{M}_k\hat{M}_\ell+\cdots \,.
	\end{align} 
The resistivity can be obtained as the matrix inverse of the conductivity tensor. This multipolar structure brings rich and unique anisotropy in the anomalous Hall effect, as demonstrated through the symmetry analysis and in real materials.  As a direct application, our theory also provides complete guiding principles for observing the in-plane anomalous Hall effect. The validity of our theory is also confirmed by existing experiments of the in-plane anomalous Hall effect in Fe, Ni\cite{Peng2024} and Fe$_3$Sn$_2$\cite{Wang2024}.
 
 Our theory can be directly generalized to two-dimensional materials. The discussion in materials without any $\tilde{C}_{nz}$ axis is straightforward. We then focus on materials with a $\tilde{C}_{nz}$ axis and discuss the possible in-plane anomalous Hall effect. The dipole cannot contribute and the octupole can in the form of the O-IV type of anisotropy. The candidate materials include CrTe$_2$\cite{Sun2020} and Cr$_2$Ge$_2$Te$_6$\cite{Fang2018} as examples.
 
 Our theory also applies in collinear antiferromagnets with the replacement of $\hat{\bm M}$ with the direction of the collinear antiferromagnetic order $\hat{\bm N}$. In our framework, this can be done by focusing on the spin order $\hat{\bm m}_1$ at a fixed lattice site in the unit cell. Since the two spins in a unit cell are correlated in a fixed manner, the Neel vector has a fixed relation with $\hat{\bm m}_1$. The multipoles such as the dipole in Eq.~\eqref{eq_dip} and the octupole in Eq.~\eqref{eq_oct} are now defined using $\hat{\bm N}$ instead of $\hat{\bm M}$. The Hamiltonian in Eq.~\eqref{eq_ham0} is still valid by replacing $\hat{\bm M}$ with $\hat{\bm N}$ in the exchange term. One can proceed with the definition of the spin frame and the resulting spin-orbit vectors. Since the spin group for collinear antiferromagnets still contains the pure spin rotation $C_{\infty z}$,  one can still transform the Taylor expansion with respect to the spin-orbit coupling to a multipolar expansion with respect to $\hat{\bm N}$. To determine the coefficients in the expansion, one should use the full spin space group symmetry instead of the point group in the ferromagnet, since the spin group in the collinear antiferromagnet is not the direct product of the spin-only group and the point group any more.

 Our theory can also extend to noncollinear antiferromagnet. In this case, one should treat the spin order as a rigid body. The spin frame should be defined in the following way: the spin at one site is chosen as reference and as the spin order changes as a whole, the $z$-axis of the spin frame follows the spin direction at that site; the spin at another site whose direction is not parallel with the first should then be chosen, and the $x$-axis is parallel with the cross product of the two spin directions. In this way, all of the three Euler angles are in use now, in sharp contrast to the ferromagnetic and collinear antiferromagnetic case, where the third Euler angle is essentially a gauge freedom. Due to the complex nature of the spin group in noncollinear antiferromagnet, the perturbation expansion of the spin-orbit coupling cannot reduce to a multipole expansion of the spin order any more, and the connection of $\bm \sigma^H$ with the spin order should be determined by the exact form of the spin-orbit vectors. The symmetry analysis of the expansion coefficient requires more complicated noncollinear spin space groups, in which the transformations of the color and spatial indices can be very different\cite{Chen2024}.

Finally, we comment that our theory provides a general framework for discussing the relation between various transport and optical phenomena and the ferromagnetic or the collinear antiferromagnetic order.  The key modification is to use the spin group analysis to identify the point-group symmetry in the color space. All the other discussion will proceed in a similar fashion with that in the anomalous Hall effect. Potential topics include the anomalous Nernst effect, the anomalous Eitinghausen effect and the magneto-optical Kerr effect.

\begin{acknowledgments}
The authors thank Stefan Bl$\mathrm{\ddot{u}}$gel and Qihang Liu for valuable discussions. The authors are supported by the National Key R${\rm \&}$D Program under grant Nos. 2022YFA1403502 and the National Natural Science Foundation of China (12234017). Z. L. is also supported by Postdoctoral Fellowship Program of CPSF (GZC20232562) and fellowship from the China Postdoctoral Science Foundation (2024M753080). Y. G. and Q. N. are also supported by the Innovation Program for Quantum Science and Technology (2021ZD0302802). D.H. is also supported by the USTC Center for Micro- and Nanoscale Research and Fabrication, National Natural Science Foundation of China (12074366), and Fundamental Research Funds for the Central Universities (WK9990000116).  The supercomputing service of USTC is gratefully acknowledged.
\end{acknowledgments}

\end{document}